\documentclass[manuscript]{aastex}

\usepackage{amsmath}
\usepackage{graphicx}
\usepackage{pdflscape}
\usepackage[hidelinks]{hyperref}
\usepackage{wrapfig}

\usepackage{float}

\slugcomment{\textbf{2024 November 15}}

\shorttitle{Non-Gravitational Forces }

\begin{document}

\title{Non-Gravitational Forces in Planetary Systems }
\author{
David Jewitt$^{1}$
} 
\affil{$^1$Department of Earth, Planetary and Space Sciences, University of California at Los Angeles, Los Angeles, CA 90095-1567}

\email{djewitt@gmail.com}

\begin{abstract}

Non-gravitational forces play surprising and, sometimes, centrally important roles in shaping the motions and properties of small planetary bodies.  In the solar system, the morphologies of comets, the delivery of meteorites and the shapes and dynamics of asteroids and binaries are all affected by non-gravitational forces. In exoplanetary systems and debris disks, non-gravitational forces affect the lifetimes of circumstellar particles and feed refractory debris to the photospheres of the central stars. Unlike the gravitational force, which is a simple function of the well known separations and masses of bodies, the non-gravitational forces are frequently functions of poorly known or even unmeasurable physical properties.   Here, we present order-of-magnitude descriptions of non-gravitational forces, with examples of their application.

\end{abstract}


\section{Introduction}
Gravity alone provides an ample description of the dynamics and of many physical properties of planetary mass bodies.  However, scientific interest is increasingly focused on smaller bodies in orbit around the Sun and other stars, and small bodies are additionally susceptible to a host of other forces.  These so-called non-gravitational forces include recoil and torque from anisotropic mass loss, radiation pressure, Poynting-Robertson drag,  Yarkovsky force, YORP torque and forces from magnetic interactions with the solar wind.  Important examples of phenomena that cannot be understood using gravity alone are numerous, ranging from the motion of dust particles in comets, the Zodiacal cloud and debris disks, to the orbital drift of asteroids and their delivery into planet-crossing orbits, to the centripetal shaping and disintegration of comets and asteroids, to the formation of asteroid binaries and pairs.  

Unfortunately, the research literature tends to present non-gravitational forces either in excruciating and largely unhelpful detail or, more usually,  without meaningful discussion of any sort.  Indeed, non-gravitational forces are often hidden as lines of code in elaborate numerical models, where their practical role is to help to improve the fit to data by adding extra degrees of freedom.  Sadly, they sometimes do so without giving a parallel improvement in our understanding of the relevant physics.

The objective of the current paper is to provide a highly simplified but nevertheless informative account of the different non-gravitational forces that are important in the solar system.  We also add some pointers to sample applications from the literature, in a style suitable for the non-specialist.
With this in mind, in the following we either neglect geometrical factors representing body shape, or approximate  the relevant bodies as spheres, with bulk density $\rho$ [kg m$^{-3}$] and radius $a$ [m].  The orbits of all bodies are assumed to be circular, so that the heliocentric distance is also the semimajor axis.  Of course, real bodies are not spherical and real orbits are not circles.  These simplifying approximations remain useful, however, by giving a guide to the order of magnitude of the forces and timescales involved. 

When the heliocentric distance, $r_H$ [meters], is expressed in au, we give it the symbol $r_{au}$.  For example, the flux of sunlight, $F_{\odot}$ [W m$^{-2}$], which plays an obvious role in several of the non-gravitational forces, is given by

\begin{equation}
F_{\odot} = \frac{L_{\odot}}{4 \pi r_H^2}~~~~~\textrm{or, equivalently,}~~~F_{\odot} = \frac{S_{\odot}}{r_{au}^2},
\end{equation}

\noindent where $L_{\odot} = 4\times10^{26}$ W is the luminosity of the Sun and $S_{\odot}$ = 1360 W m$^{-2}$ is the solar constant.

\section{Sublimation Recoil}
\label{recoil}

By far the largest non-gravitational force considered here is that due to the sublimation of ice from small bodies heated by the Sun.  Sublimated volatiles freely expand into the near vacuum of interplanetary space, carrying with them momentum and exerting a recoil force on the ice-containing parent body.  Since sublimation is exponentially dependent on temperature, most sublimation occurs on the hot day side of the nucleus and the resulting recoil force is primarily anti-solar.  This sublimation recoil force is $\mathcal{F} = k_R \dot{M} V_{th}$, where $\dot{M}$ [kg s$^{-1}$] is the sublimation rate, $V_{th}$ [m s$^{-1}$] is the bulk speed of the outflowing gas and $k_R$ is a dimensionless constant that describes the angular dependence of the flow. Purely collimated flow would have $k_R$ = 1, while isotropic outgassing would have $k_R$ = 0, meaning no net force on the nucleus. The best estimate of $k_R$ is based on measurements from the unusually well-studied comet 67P/Churyumov–Gerasimenko, for which $k_R \sim$ 1/2 \citep{Jew20}.  This is also the value expected from uniform sublimation across the sunward facing hemisphere of a spherical nucleus.

We can also write $\dot{M} = \mu m_H Q_g$, where $\mu$ is the molecular weight of the sublimated ice, $m_H = 1.67\times10^{-27}$ kg is the mass of the hydrogen atom and $Q_g$ is the gas production rate in molecules per second.  Setting $\mathcal{F} = M \alpha_S$, where $M = 4\pi \rho a^3/3$ is the body mass, gives a non-gravitational acceleration (NGA)

\begin{equation}
\alpha_S = \frac{3k_R \mu m_H}{4\pi \rho a^3}  Q_g V_{th}
\label{alpha}
\end{equation}

\noindent for a spherical body of radius, $a$, and density $\rho$.

The outflow speed of the gas, $V_{th}$, is roughly equal to the thermal speed of the constituent molecules

\begin{equation}
V_{th} = \left(\frac{8 k_B T}{\pi \mu m_H}\right)^{1/2}.
\label{gas_speed}
\end{equation}

\noindent Here, $k_B = 1.38\times10^{-23}$ J K$^{-1}$ is Boltzmann's constant, and temperature $T$ refers to the surface temperature of the sublimating surface. This is typically depressed below the local radiative equilibrium temperature because a fraction of the absorbed solar energy that would otherwise drive radiative cooling is instead used to break bonds between molecules in the process of sublimation.  While $T \propto r_H^{-1/2}$ in radiative equilibrium, the temperature of sublimating ice is an even weaker function of heliocentric distance because of this depression.  We take $V_{th}$ = 500 m s$^{-1}$ at 1 au and assume for simplicity that this speed applies across the terrestrial planet region.

The  temperature and sublimation rate per unit area, $f_s(T)$ [kg m$^{-2}$ s$^{-1}$],  are connected by the energy balance equation for a surface element of area oriented with its normal offset from the direction of illumination by angle $\theta$;

\begin{equation}
\frac{S_{\odot}}{r_{au}^2} (1-A) \cos\theta  = \varepsilon \sigma T^4 + H(T) f_s(T).
\label{sublimation}
\end{equation}


\noindent Here, $A$ and $\varepsilon$ are the Bond albedo and emissivity of the surface, $\sigma = 5.67\times10^{-8}$ W m$^{-2}$ K$^{-4}$ is the Stefan-Boltzmann constant, and $H(T)$ [J kg$^{-1}$] is the latent heat of sublimation for the ice in question.  An additional, generally small term for heat conducted beneath the surface has been neglected from Equation \ref{sublimation}.

Additional information is needed to solve Equation \ref{sublimation}.  The temperature dependence of $f_s$ can be obtained from the Clausius-Clapeyron equation for the slope of the solid/vapor phase boundary (the relation $dP/dT = P H/(N_A k T^2)$, where $N_A$ is Avagadro's number, and $k$ is the Boltzmann constant is only applicable when $H$ is independent of $T$) or, better, from laboratory measurements of the sublimation vapor pressure over ice as a function of temperature.  The optical properties $A$ and $\varepsilon$ have a minor effect on the solution provided $A \ll $1 and $\varepsilon \gg$ 0.  In practice, $A$ = 0 and $\varepsilon$ = 0.9 are widely assumed.

\begin{figure}
\epsscale{0.99}
\plotone{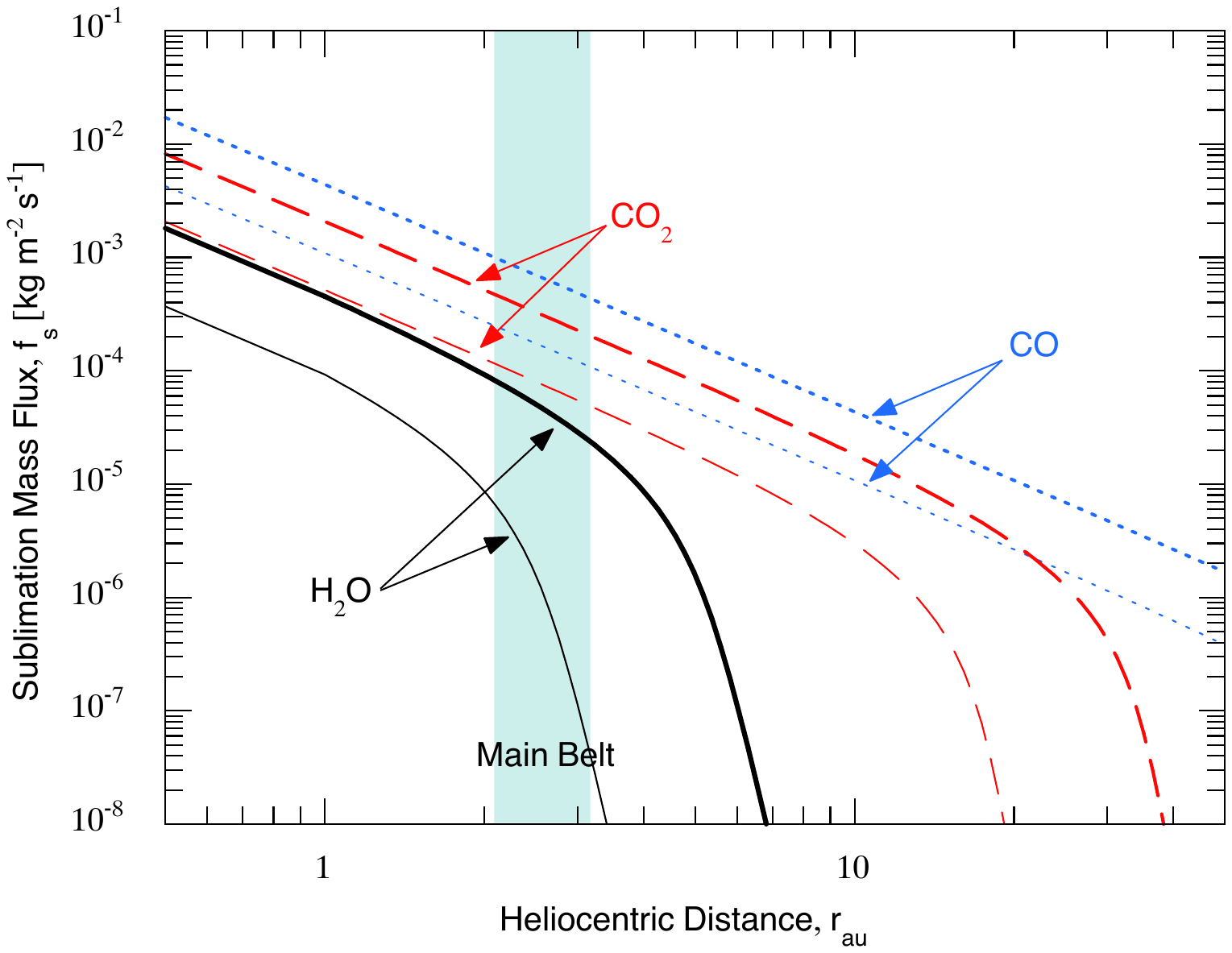}
\caption{Equilibrium sublimation mass fluxes as a function of heliocentric distance for H$_2$O (solid black lines), CO$_2$ (long-dashed red lines) and CO (short-dashed blue lines) ices, computed from Equation \ref{sublimation}.  Two models are shown for each ice.  The upper model  for each ice shows  sublimation at the subsolar temperature, taken as the highest temperature on a spherical body, while the lower model shows sublimation at the local isothermal backbody temperature, which is the lowest possible temperature.  The location of the asteroid belt is marked as a shaded blue rectangle, with inner and outer edges at 2.1 and 3.2 au.
 \label{ices}}
\end{figure}

In Figure \ref{ices} we show solutions to Equation \ref{sublimation}  for the three most abundant cometary ices of water, carbon dioxide and carbon monoxide.  These three, having approximate latent heats $H$ = 2.8$\times10^6$ J kg$^{-1}$, 0.57$\times10^6$ J kg$^{-1}$ and 0.29$\times10^6$ J kg$^{-1}$, respectively, are representative of low, medium and high volatility solids.  

All three plotted curves trend asymptotically towards $f_s \propto r_{au}^{-2}$ as $r_{au} \rightarrow$ 0.  This is because the exponential temperature dependence of $f_s$ is stronger than $T^4$, so that the second term on the right of Equation \ref{sublimation} dominates the first at the high temperatures found at small $r_{au}$.  Setting the radiative term in Equation \ref{sublimation} equal to zero gives

\begin{equation}
f_s \sim \frac{S_{\odot}}{H(T) r_{au}^2} 
\label{sublimation2}
\end{equation}

\noindent where we have set $A$ = 0 and $\theta$ = 0 for simplicity.   Equation \ref{sublimation2} gives a useful approximation even at $r_{au}$ = 1 au, where $f_s = 6\times10^{-4}, 2.4\times10^{-3}$, and $4.7\times10^{-3}$ kg m$^{-2}$ s$^{-1}$, respectively, for water, carbon dioxide and carbon monoxide ices on a flat plate oriented normal to the Sun.  Ices like carbon monoxide are so volatile that the $f_s \propto r_{au}^{-2}$ regime extends over the entire planetary region while, for the intermediate volatile carbon dioxide ice, the sublimation rate inflects closer to the Sun, but still beyond the orbit of Jupiter.

Across the $\sim$1 au width of the asteroid belt, shown in Figure \ref{ices} as a blue shaded region, the sublimation flux of water varies appreciably, by a factor of $\sim$4 in the high temperature limit and by a factor $\sim$ 200 in the low temperature limit.  This strong variation explains the sudden rise of the outgassing activity observed in comets as they cross asteroid belt distances (e.g., \cite{Biv02}) and the strong concentration of outgassing from main-belt comets near perihelion \citep{Hsi15}.

The total mass loss rate from a sublimating body is given by the integral $\dot{M} = \int f_s dS$  over the surface, $S$, taking account of the object shape because it controls the local solar incidence angle $\theta$.  In practice, evaluation of this integral is impossible because the shapes of most bodies are not well determined and the distribution of surface ice is patchy.   Moreover, ice can sublimate from beneath the  surface, resulting in a complicated and generally underconstrained thermophysical problem for which simple solutions do not exist.  Therefore, while it is useful to consider the broad trends shown in Figure \ref{ices}, it is not in general possible to calculate a global value of $f_s$ or $\dot{M}$ from first principles.

\subsection{Examples}

\begin{itemize}
\item \textbf{Comets:} Non-gravitational motions are especially obvious in the orbits of short-period comets, where the cumulative effects can be measured over many orbits.  The non-gravitational acceleration is conventionally written \citep{Mar73}

\begin{equation}
\alpha_{S} = g(r_H) (A_1^2 + A_2^2 + A_3^2)^{1/2}
\end{equation}

\noindent in which $A_1$, $A_2$ and $A_3$ are the components of the non-gravitational acceleration in the directions radial to the Sun, perpendicular to $A_1$ but in the orbital plane, and perpendicular to the orbital plane, respectively.  The acceleration components are conventionally expressed in au day$^{-2}$, where 1 au day$^{-2}$ = 20.1 m s$^{-2}$.  For comets, $g(r_H)$ conventionally takes a contrived analytic form based on solutions to Equation \ref{sublimation} for water ice \citep{Mar73}.  For asteroids, $g(r_H) = 1/r_{au}^2$, where $r_{au}$ is the heliocentric distance in au.  

\begin{figure}
\epsscale{0.89}
\plotone{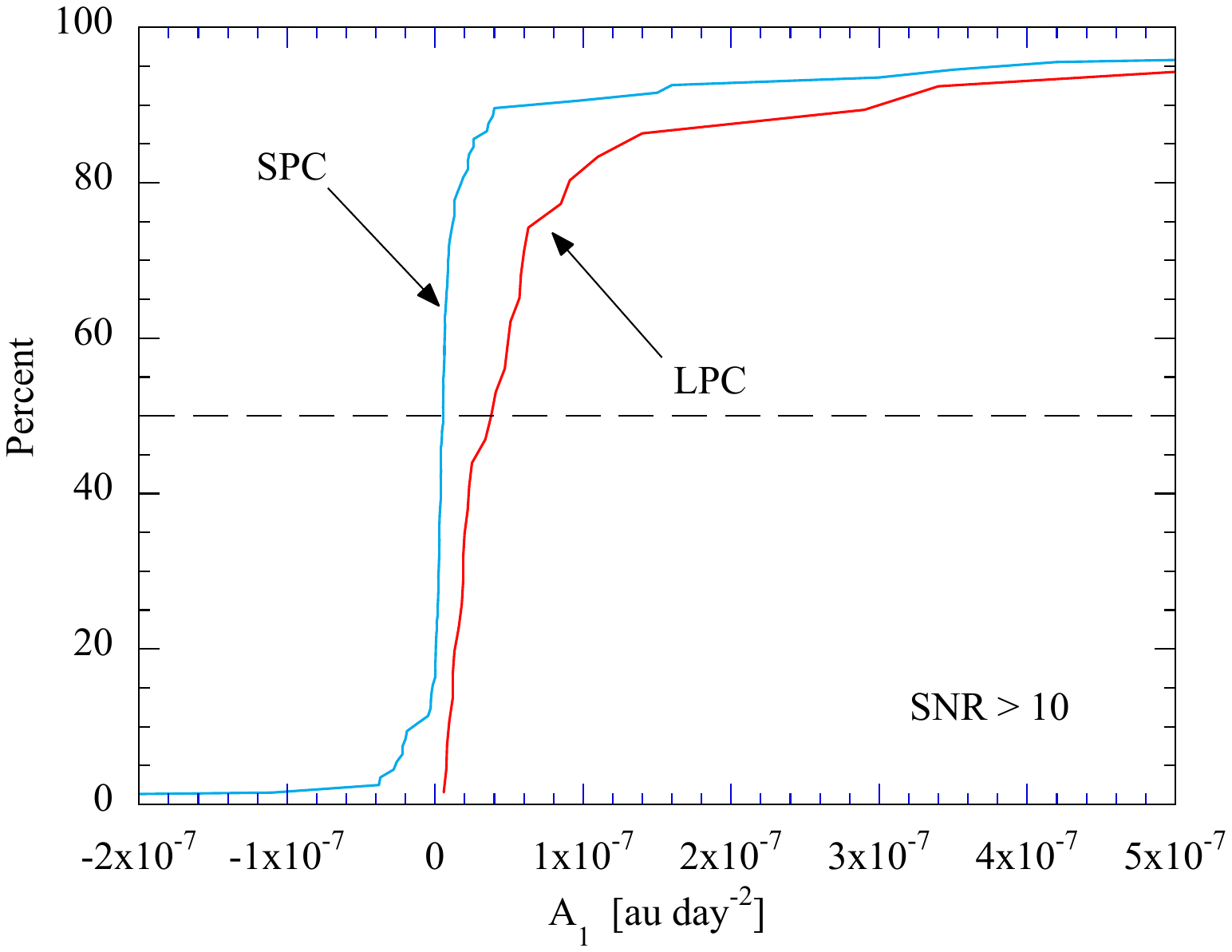}
\caption{Cumulative distribution of the $A_1$ (radial) component of the non-gravitational acceleration for 101 short-period comets (SPCs: blue line, 2 $\le T_J \le$ 3) and 33 long-period comets (LPCs: red line, $T_J <$ 2).  Only comets with $A_1$ measured to $>10\sigma$ significance are plotted. The LPCs show systematically larger $A_1$, consistent with having smaller and/or less dense nuclei, and with having larger outgassing rates per unit area.}
\label{A1_plot}
\end{figure}

The cumulative distributions of $A_1$ for Jupiter family comets and long-period comets are shown separately in Figure \ref{A1_plot} using NGA parameters from JPL Horizons\footnote{https://ssd.jpl.nasa.gov/}.  The astrometric measurements from which NGA is determined are fraught with systematic errors (for example, in a comet with a coma, the center of light and the nucleus may not coincide) and some measure of judgement is required in the rejection of unreliable data in determining the likely best fit.  To try to reduce this problem, we have plotted in Figure \ref{A1_plot} only measurements having signal-to-noise ratio $>$ 10.  The figure shows that the $\alpha_s$ distributions are significantly different for LPCs and SPCs, with the LPCs showing a larger acceleration (rms value $16\times10^{-7}$ au day$^{-2}$, corresponding to NGA $\alpha = 32\times10^{-6}$ m s$^{-2}$) than the SPCs (rms $4\times10^{-7}$ au day$^{-2}$ or NGA $\alpha = 8\times10^{-6}$ m s$^{-2}$).  This difference could indicate a systematic difference in the sizes or densities of SPC and LPC nuclei, or a difference in the outgassing rates per unit area of surface, or some combination of these effects.  A more prosaic explanation is a measurement bias;  small NGAs cannot generally be accurately measured in LPCs because they are observed at only one apparition, whereas SPCs can be measured over multiple orbits and longer periods of time.  Some 16/101 SPCs (but 0/33 LPCs) exhibit negative $A_1$ parameters, which indicate unexpected sunward acceleration.  These determinations, which might indicate that the astrometric errors are underestimated, cannot be explained by  recoil from dayside sublimation.

\item \textbf{Dark Comets:} 
Some small, apparently inert bodies exhibit substantial NGA, yet present no evidence for outgassing. Notable examples include the interstellar interloper 1I/’Oumuamua \citep{Mic18} and the $\sim$150 m radius asteroid 523599 (2003 RM) \citep{Far23}, as well as a number of tiny asteroids (essentially, boulders) in the
3 to 15 m radius range \citep{Sel23}. The magnitude of the NGA in these evocatively named “Dark Comets” is too large to reflect the action of radiation pressure or Yarkovsky force but consistent with small sublimation rates.  
For example, the measured acceleration of 2003 RM is $\alpha = 2\times10^{-12}$ au day$^{-2}$ ($\alpha = 4\times10^{-11}$ m s$^{-2}$).  From Equation \ref{alpha} with $k_R$ = 0.5, $\mu$ =18 (for H$_2$O), $\rho = 10^3$ kg m$^{-3}$ and $V_{th}$ = 500 m s$^{-1}$, we find $Q_g = 4\times10^{25}$ s$^{-1}$ (equivalent to 1 kg s$^{-1}$), which might be small enough to have escaped detection.   The Tisserand parameter with respect to Jupiter of 2003 RM is formally that of a comet ($T_J$ = 2.96) in which case the presence of ice and outgassing should not be surprising. However, Farnocchia et al. (2023) prefer an origin in the outer asteroid belt and derive a much smaller $Q_g \sim 10^{23}$ s$^{-1}$.

The  boulders described by \cite{Sel23} can be accelerated by sublimation at even smaller rates.  
For example, the strongest (8$\sigma$) detection of NGA is in the 4 m radius boulder 2010 RF12, with $A_3 = (-0.17 \pm 0.02)\times10^{-10}$ au day$^{-2}$ ($\alpha_S$ = 3.3$\times10^{-10}$ m s$^{-2}$).  Substitution into Equation \ref{alpha} gives $Q_g \sim 10^{19}$ s$^{-1}$ (roughly $3\times10^{-7}$ kg s$^{-1}$) which is small enough to have escaped detection by any existing direct technique.  Radiation pressure acceleration should be of the same order as $\alpha_s$ but cannot account for $A_3$, which acts perpedicular to the projected radial line.  The main puzzle  presented by the accelerated boulders is their implied short mass-loss lifetimes. 2020 RF12, with mass $\sim3\times10^5$ kg, can sustain mass loss at $3\times10^{-7}$ kg s$^{-1}$ for only $10^{12}$ s (3$\times10^4$ years).  The conduction timescale is even shorter ($\sim$6 months for a compact rock with diffusivity $\kappa = 10^{-6}$ m$^2$ s$^{-1}$ and $\sim$500 years if it is a porous dust ball with the very low diffusivity $\kappa = 10^{-9}$ m$^2$ s$^{-1}$; See Appendix A).  As a result, the internal temperatures of this and other boulders near 1 au would quickly equilibrate to orbit-averaged values that are too high ($\sim$300 K) for water ice to survive.  Even if the mass loss is intermittent, it is hard to see how such tiny boulders could retain ice on dynamically relevant (Myr) timescales.
\end{itemize}

\section{Radiation Pressure}

A radiant energy flux density  $F_{\nu}$ [J m$^{-2}$ s$^{-1}$ Hz$^{-1}$] corresponds to a photon flux $F_{\nu}/(h\nu)$ [photon m$^{-2}$ s$^{-1}$ Hz$^{-1}$], where $h = 6.63\times10^{-34}$ J s is Planck's constant and $h \nu$ is the energy of a photon having frequency $\nu$. The momentum of a single photon is $h/\lambda = h \nu/c$, where $\lambda$ is the wavelength, and $c = 3\times10^8$ m s$^{-1}$ is the speed of light.  Then, considering the Sun as the source of photons, the  flux of momentum in photons of  frequency $\nu \rightarrow \nu + d\nu$ is $dP_{r,\nu} = F_{\nu}/(h\nu)\times (h\nu/c) d\nu$.  When integrated over all frequencies this gives a pressure

\begin{equation}
P_r = \frac{F_{\odot}}{c}~~\textrm{[N m$^{-2}$]}
\label{Pr}
\end{equation}

\noindent in which $F_{\odot} = \int_0^{\infty} F_{\nu} d\nu$. Equation \ref{Pr} is the radiation pressure.  For example, at $r_H$ = 1 au, where the flux of sunlight is given by the solar constant, $S_{\odot} = 1360$ W m$^{-2}$, Equation \ref{Pr} gives $P_r =4.5\times10^{-6}$ N m$^{-2}$ (about 4 million times less than the pressure exerted by the weight of a sheet of paper). This tiny pressure is about 10$^5$ times smaller than the  pressure due to water ice sublimation at 1 au, and is insignificant on macroscopic bodies, but can dominate the motion of small particles.

The force exerted by radiation impinging on a spherical grain of radius $a$, is $\mathcal{F} = Q_{pr} \pi a^2 F_{\odot}/c$, where $Q_{pr}$ is a size and wavelength dependent dimensionless multiplier\footnote{$Q_{pr}$ is the ratio of the effective cross-section for radiation pressure to the geometric cross-section of the particle.  It is a function of the composition, shape, structure and size of a particle relative to $\lambda$, the wavelength of radiation with which it interacts.  The limits are $Q_{pr} \ll$ 1 as $a/\lambda \rightarrow$ 0 while $Q_{pr} \rightarrow$ constant as $a/\lambda \rightarrow \infty$.  In between, $Q_{pr}$ varies with $a/\lambda$ in a complicated way, reflecting interactions between electromagnetic waves as they pass, and pass through, the particle (\cite{Bur79}, \cite{Boh83}).  In planetary science and astronomy, the zeroth order approximation is to set $Q_{pr}$ = 0 for $a < \lambda$ and $Q_{pr}$ = 1 for $a \ge \lambda$.  For many natural particle size distributions in which the smallest particles are the most abundant, this approximation gives rise to the rule-of-thumb that observations primarily sample particles with $a \sim \lambda$.   Calculation of $Q_{pr}$ for homogeneous spheres uses the Mie Theory. For other shapes and for porous and fractal particles of relevance to natural solar system particles, there is no analytic theory and $Q_{pr}$ must be calculated numerically (e.g., \cite{Sil16}).} and $F_{\odot}$ [W m$^{-2}$] is the local flux of sunlight.   The radiation pressure acceleration is proportional to the cross-section per unit mass of the accelerated particle, and so is inversely related to the particle size.  For a spherical particle of density $\rho$, this results in an acceleration

\begin{equation}
\alpha_{rad} = \frac{3 Q_{pr} F_{\odot}}{4\rho c a}.
\label{rp1}
\end{equation}

\noindent  The inverse dependence shows that small, low density particles can be more strongly accelerated by radiation than large, high density particles.   In the case of the radiation field around the Sun, we note that the flux is given by $F = L_{\odot}/(4\pi r_H^2)$, where $L_{\odot} = 4\times10^{26}$ W is the solar luminosity and $r_H$ is the heliocentric distance in meters.  Substituting gives

\begin{equation}
\alpha_{rad} = \frac{3 Q_{pr} }{4 \rho c a } \left(\frac{L_{\odot}}{4 \pi r_H^2}\right)
\label{rp}
\end{equation}

\noindent The numerical multiplier in Equations \ref{rp1} and \ref{rp} is specific to the assumed spherical particle shape.  Nevertheless, the equations give a useful approximation to the acceleration induced by radiation pressure as a function of particle size, and density.  It is useful to normalize $\alpha_{rad}$ by the local solar gravitational acceleration 

\begin{equation}
g_{\odot} = \frac{G M_{\odot}}{r_H^2}.
\label{gsun}
\end{equation}

\noindent The ratio $\beta_{rad} = \alpha_{rad}/g_{\odot}$ is given by

\begin{equation}
\beta_{rad} = \frac{3 Q_{pr}}{16 \pi \rho c a } \left(\frac{L_{\odot}}{G M_{\odot}}\right)
\label{beta}
\end{equation}

\noindent which is a function of both particle  ($\rho$, $a$, $Q_{pr}$) and stellar ($L_{\odot}$, $M_{\odot}$) properties.  

For example, consider spherical dust grains with $\rho$ = 1000 kg m$^{-3}$ and $Q_{pr}$ = 1.  Substitution  into Equation \ref{beta} gives 

\begin{equation}
\beta_{rad} = \frac{0.6}{a_{\mu}}, 
\label{beta2}
\end{equation}

\noindent where $a_{\mu}$ is the radius expressed in microns.   $\beta_{rad}$ is independent of heliocentric distance because gravity and radiation pressure both vary inversely with the square of $r_H$.    Equation \ref{beta2} is a useful guide for homogeneous spheres, but it is important to note that the $\beta$ parameter for other particle shapes cannot be so simply estimated because $Q_{pr}$ can assume very different values.  Likewise, $\beta_{rad} \gg$ 1 is never encountered, regardless of how small the particles are \citep{Sil16}.

\subsection{Examples:}

\begin{itemize}

\item \textbf{{Comet Tails:}}
\begin{figure}
\epsscale{0.99}
\plotone{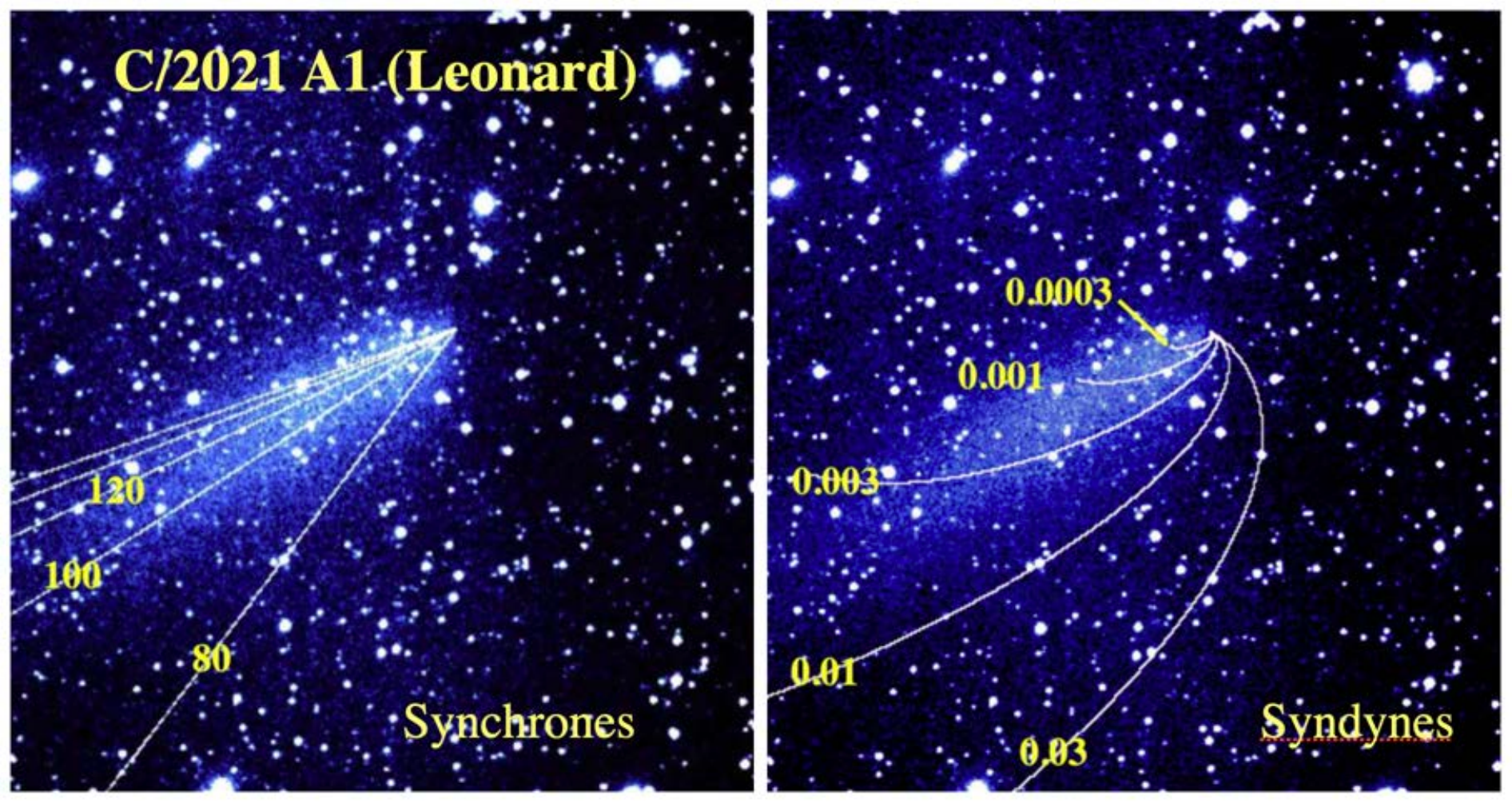}
\caption{Disintegrated long-period comet C/2021 A1 (Leonard) on UT 2022 March 31 when at $r_H$ = 1.756 au outbound.  Left panel shows synchrones for particles ejected from 80 to 160 days before the date of observation.  Right panel shows syndynes for particles with $\beta$ = 0.0003, 0.001, 0.003, 0.01, and 0.03, as marked.  The linear morphology of the tail is better matched by the straight synchrone models than by the curved syndynes. Ejection occurred 110$\pm$10 days before the image was taken, i.e., on UT 2021 December 11$\pm$10.  Calculation by Yoonyoung Kim in  \cite{Jew23}.
 \label{synsyn}}
\end{figure}
Radiation pressure acceleration has been detected in  bodies as large as the $\sim$4 to 10 m diameter asteroid 2011 MD, for which $\beta_{rad} \sim 10^{-7}$ (\cite{Mic14}; \cite{Mom14}) but the classic application is to the motion of small particles in comets.  The dust particle trajectory after release from a comet nucleus is determined by the ejection velocity and $\beta_{rad}$. In the  theory of comet tails  by  \cite{Fin68}, dust particles are assumed to be released from the nucleus  with zero relative speed and then to be accelerated by solar gravity and radiation pressure.  Two limiting cases are often considered; synchrones show the locus of positions of dust particles having a range of sizes but released from the nucleus at the same time, while syndynes show the locus of positions of particles of one size (i.e., one $\beta_{rad}$) but released over a range of times.  Synchrones project on the sky as straight lines, with a position angle that depends on the time of ejection.  Syndynes are curved by the orbital motions of the particles.   See Figure \ref{synsyn} for examples.

 Syndyne/Synchrone analysis enables a simple assessment of cometary dust properties. For example, depending on the observing geometry, a linear tail morphology may suggest a synchrone-like structure indicative of impulsive ejection.  Linear tails are observed in outbursting comets and in dust released by energetic impacts between asteroids  \citep{Jew12}, where the position angle of the tail gives a useful estimate of the ejection date.  On the other hand, broad, fan-like tail structures are indicative of protracted emission for which syndyne models give constraints on the particle sizes.  When observed from a position close to the comet orbital plane, however, syndynes and synchrones overlap so strongly that their diagnostic power is largely lost.  
 
 With the aid of fast computers, more sophisticated synthetic models of comets can be used to explore a wide range of dust and ejection conditions.  Such Monte Carlo models involve many parameters (including the form of the size distribution (usually assumed to be a power law), the largest and smallest particle radii, their ejection velocities as a function of particle size, the angular distribution of the ejection from the nucleus and its time dependence) and so are non-unique.  Nevertheless, Monte Carlo coma models are useful in narrowing the range of allowed parameters needed in order to match a given observation. Much of what we know about the solid particles ejected from comets has been determined this way.  In the example of Figure \ref{synsyn}, inferences from the syndynes and synchrones were supported by a full Monte Carlo simulation.  A 0.6 km radius nucleus disintegrated suddenly into a broad distribution of particle sizes resembling a -3.5 index power law, with no fragments larger than 60 m.

\item \textbf{{$\beta$ Meteoroids:}} The radiation pressure force opposes gravity such that, at heliocentric distance $r_H$, the net sunward attraction is $(1-\beta_{rad})G M_{\odot}/r_H^2$ and the effective escape velocity is $V_e = (2 (1-\beta_{rad})G M_{\odot}/r_H)^{1/2}$.  Comparing this to the local Keplerian speed, $V_K = (G M_{\odot}/r_H)^{1/2}$ shows that particles in circular orbits with $\beta_{rad} >$ 1/2 ($a \lesssim$ 1 $\mu$m by Equation \ref{beta2}) should be gravitationally unbound and promptly leave the solar system along nearly radial trajectories\footnote{For a non-circular orbit, larger particles with $\beta_{rad} > (1-e)/2$ can be ejected by radiation pressure.}. Such particles, called $\beta$-meteoroids \citep{Zoo75}, have been detected.  They are sub-micron sized dust particles that stream away from the Sun faster than the gravitational escape speed under the action of radiation pressure. The $\beta$  particles are produced  when larger dust particles sublimate and shrink upon approach to the Sun, as well as by collisional shattering of larger grains from the Zodiacal cloud that are spiraling in under the action of Poynting-Robertson drag (\cite{Man21}; see section \ref{PRdrag}). Their source region is concentrated at 10 to 20 R$_{\odot}$ (0.05 to 0.1 au), where blackbody temperatures lie in the range 880 K to 1240 K and both collision and sublimation rates are high \citep{Sza21}. 
The effective mass loss rate in $\beta$ meteoroids is estimated between $\gtrsim10^2$ kg s$^{-1}$ \citep{Sza21} and 10$^3$ kg s$^{-1}$ \citep{Gru85}. If sustained over the age of the solar system, a dust mass $\sim10^{19}$ to 10$^{20}$ kg must have been expelled by radiation pressure as $\beta$ meteoroids, corresponding to a few percent of the current mass in the asteroid belt.  

\item \textbf{{Kuiper Belt Dust:}}
While $\beta$ meteoroids are best known from spacecraft measurements obtained in the inner and middle solar system, more distant counterparts may originate as well in dust sources in the outer solar system. In this regard, recent impact counter measurements from the New Horizons spacecraft have been used to suggest (albeit at only 2$\sigma$ significance) an excess concentration of dust grains (radius $\sim$0.6 $\mu$m, $\beta \sim$ 1) beyond the 47 au edge of the classical Kuiper belt \citep{Don24}. These dust grains might  be produced by collisional shattering in the main Kuiper belt followed by radiation pressure acceleration outwards as Kuiper belt $\beta$ meteoroids.  Meanwhile, $\beta$ meteoroids ejected from other stars presumably contribute to the flux of interstellar dust particles entering the solar system \citep{Lan00}.

\item \textbf{{Extended Debris Disk Dust:}}  Many stars are encircled by disks in which the dust lifetime is shorter than the main-sequence age of the star.  In these so-called Debris Disks, the dust must be replenished, presumably by collisional destruction of unseen parent bodies in what are, effectively, extrasolar Kuiper belts  \citep{Hug18}.  Since the luminosity of a main-sequence star scales in proportion to a high power of its mass (e.g., $L_{\star} \propto M_{\star}^{x}$ with $x$ = 3.5 to 4), the importance of radiation pressure should grow with stellar mass.  A classic example is provided by the disk of $\alpha$ Lyrae (Vega), a $\sim$700 Myr old A0V star 2.2 $M_{\odot}$ in mass with luminosity 47 $L_{\odot}$.  All else being equal, $\beta_{rad}$ is $\sim$20 times larger at $\alpha$ Lyrae than around the Sun, leading to the immediate expulsion of particles 20 times larger (i.e., 10s of $\mu$m instead of $\sim$1 $\mu$m).  The large extent of the $\alpha$ Lyrae dust (which reaches $\sim$1000 au  from the star) is likely due to the action of radiation pressure \citep{Su05}.  The implied dust mass flux ($\gtrsim 10^{12}$ kg s$^{-1}$) is too large to be sustained over the life of Vega.  \cite{Su05} suggest that the dust could be the recent product of a massive collision in the disk.

\item \textbf{{Lifting of Dust:}}
Dust can be lifted from the surface of a small body of mass $M_a$ and radius $r$ when the solar radiation pressure acceleration, $\beta_{rad} G M_{\odot}/r_H^2$ exceeds the local gravitational attraction to the body, $GM_a/r^2$.  The critical particle size for ejection obtained by setting these accelerations equal is

\begin{equation}
a_c \sim 10 \left(\frac{1~\textrm{km}}{r}\right) \left(\frac{1}{r_{au}}\right)^2
\end{equation}

\noindent where $a_c$ is in $\mu$m, $r_{au}$ in au and $r$ in km \citep{Jew12}.  For example, particles on a 1 km radius body  located at 1 au can be swept provided $a_c \sim 10$ $\mu$m, or smaller.  Close to the Sun (e.g., at 0.14 au, the perihelion distance of (3200) Phaethon), much larger particles ($a_c \lesssim 1$ mm) can be expelled.  \cite{Bac21} note that thermal radiation pressure from a sufficiently hot surface might also expel particles, although no established examples of this process currently exist.  Both loss processes are highly directional (e.g., solar radiation pressure will force particles back into the surface except near the terminator) and, whether by solar or thermal radiation pressure, particle ejection must overcome the cohesive forces which bind small particles to the surface.

\end{itemize}

\section{Poynting-Robertson Drag}
\label{PRdrag}
Formally a relativistic effect with a rather involved derivation \citep{Rob37}, the Poynting-Robertson drag is more simply described as a consequence of aberration.  As seen from a body moving in a circular orbit, the direction from which sunlight travels is aberrated relative to the radial direction from the Sun by an angle $\theta = \tan^{-1}(V_K/c)$, where $V_K = (G M_{\odot}/r_H)^{1/2}$ is the orbital speed and $c$ is the speed of light.  For  solar system objects of interest, $V_K \ll c$ and we may approximate $\theta \sim V_K/c$.  For example,  the Earth ($r_H$ = 1 au) has orbital speed $V_K$ = 30 km s$^{-1}$ giving $V_K/c \sim 10^{-4}$ (about 20\arcsec). 
This tiny angle gives a non-radial component of the radiation pressure force, which acts steadily against the direction of motion, and so can do work against the orbit. The result is a drag force (the Poynting-Robertson drag force) that results in  inexorable orbital decay.

To estimate the timescale for this orbital decay, $\tau_{PR}$, we assume a spherical body of radius $a$ moving in a circular orbit and use Equation \ref{rp} to write the Poynting-Robertson acceleration tangential to the orbit as 

\begin{equation}
\alpha_{PR} = \frac{3 Q_{pr} }{4\rho c a} \left( \frac{L_{\odot}}{4 \pi r_H^2}\right) \left( \frac{V_K}{c}\right).
\end{equation}

\noindent Then, the timescale for this acceleration to collapse the orbit is $\tau_{PR} \sim V_K/\alpha_{PR}$ or

\begin{equation}
\tau_{PR} \sim \frac{4 \rho a c^2}{3 Q_{pr}} \left(\frac{4\pi r_H^2}{L_{\odot}}\right).
\label{taupr}
\end{equation}

\noindent For example, the timescale for a  1 mm radius particle of density $\rho$ = 1000 kg m$^{-3}$ to spiral into the Sun from Earth orbit ($r_H$ = 1 au) under the action of the Poynting-Robertson drag is $\tau_{PR} \sim 8\times10^{13}$ s, or about 3 Myr (assuming $Q_{pr}$ = 1).  Setting $\tau_{PR} =$ 4.5 Gyr in Equation \ref{taupr} gives $a \sim$ 1.5 m: all primordial material smaller than a few meters in size is removed by Poynting-Robertson drag if it survives that long against other destructive processes (e.g., impact destruction, gravitational scattering).   

\subsection{Examples:}

\begin{itemize}
\item \textbf{{Zodiacal Dust:}}
Interplanetary dust is released by sublimating comets and, to a lesser extent, by collisions between asteroids, forming  a diffuse inner solar system structure known as the Zodiacal cloud.   Dust particles in the Zodiacal cloud have a range of sizes with a mass-weighted mean near $\bar{s}$ = 100 to 200 $\mu$m, for which the Poynting-Robertson time is $\tau_{pr} \sim$ 0.5 Myr at 1 au.  Smaller particles are quickly depleted by Poynting-Robertson drag while the survival of larger particles is limited more by collisional shattering.  The total rate of production required to maintain the Zodiacal cloud in steady state is  $\sim$10$^3$ to 10$^4$ kg s$^{-1}$, albeit with order of magnitude uncertainty (\cite{Nes11}, \cite{Rig22}, \cite{Pok24}).  The steady Poynting-Robertson rain of Zodiacal dust towards the Sun is reversed when a fraction of those particles shrink small enough (through sublimation, and/or through collisional shattering) to be expelled by radiation pressure as $\beta$ meteoroids.

\item \textbf{{Interplanetary Dust Particles:}}
Zodiacal dust particles enter the Earth's atmosphere at an average rate $\sim$1 kg s$^{-1}$ \citep{Lov93}. Entering particles smaller than a few tens of microns are decelerated by friction so high in the atmosphere that they do not melt and, when collected from the lower stratosphere, are known as Interplanetary dust particles (IDPs). The rate of IDP delivery presumably varies on long timescales as asteroids collide and produce variable amounts of  dust, which is then cleared in part by Poynting-Robertson drag.  However, stratigraphic measurements of $^3$He, which is delivered to Earth by IDPs, show only factor-of-two variations in the rate in the last 100 Myr \citep{Far21}.

On the way from their source to the Earth, the IDPs are impacted by energetic solar and cosmic ray nuclei, causing damage tracks and the transmutation of elements. \cite{Kel22} used particle track densities to calculate the effective space exposure ages of 10 $\mu$m sized IDPs. They found that about 25\% of IDPs show high cosmic ray track densities that are indicative of exposure ages $>10^6$ year. By contrast, the Poynting-Robertson lifetime of 10 $\mu$m particles at 1 au is only about 15 kyr (from Equation \ref{taupr}). The two orders of magnitude age discrepancy could indicate that these high track density IDPs originate at much larger distances, because  $\tau_{pr} \propto r_H^2$ (Equation \ref{taupr}). For example, 10 $\mu$m particles released at $r_H \gtrsim$ 10 au would have Poynting-Robertson timescales $\tau_{pr} > 10^6$ year, consistent with those measured.  The only known substantial dust source beyond 10 au is the Kuiper belt, which \cite{Kel22} propose is the source of the heavily cosmic ray damaged dust.  Independent IDP exposure age estimates based on the production of unstable Al$^{26}$ and Be$^{10}$ nuclei
lead to qualitatively consistent conclusions: IDPs originate over a vast range of heliocentric distances, with some from the Kuiper belt \citep{Fei24}.

The reported collisional production rate of dust in the Kuiper belt  ($\sim10^4$ kg s$^{-1}$)  exceeds the $\sim10^3$ kg s$^{-1}$ released from the short period comets and $\lesssim50$ kg s$^{-1}$ from long period comets \citep{Pop16}.  For this reason it should not be surprising to find an IDP contribution from the Kuiper belt.  Indeed, early  simulations of dust transport to the inner solar system suggested a maximum 25\% contribution \citep{Mor03}, consistent with the new estimate based on track densities in the IDPs.  

\item \textbf{{White Dwarf Contamination:}}
White dwarfs are degenerate post-main sequence, roughly solar mass stars collapsed to Earth-like dimensions.  Some white dwarfs exhibit excess infrared emission from circumstellar disks having characteristic radii $\lesssim$1 R$_{\odot}$.  Refractory material falling from these disks may enter the Roche sphere of the star, become tidally shredded, and then sublimate to form a gaseous metal disk which  contaminates the photosphere with rock-forming elements. Poynting-Robertson drag moves material inward to the sublimation radius to feed the metal gas disk. A straightforward application of  Equation \ref{taupr} gives timescales too long, and peak accretion rates ($\sim10$ kg s$^{-1}$) orders of magnitude too small to account for the observed degree of white dwarf pollution \citep{Far10}.  However, \cite{Raf11} showed that the effective $\tau_{PR}$ is much shorter, and the mass delivery rate much greater ($\sim$10$^5$ kg s$^{-1}$), because Poynting-Robertson drag need only move particles across a thin transition zone at the inner edge of the refractory disk into the sublimation radius.

\end{itemize}

\section{Dissipative Forces}
\subsection{Tidal Dissipation}
\label{tides}
The shapes of bodies in mutual orbit are cyclically deformed by varying gravitational forces, resulting in the dissipation of rotational and orbital energy as heat.  Over time, this dissipation can have  important dynamical consequences.  Although gravity is the driver, the fundamental origin of  energy loss  is non-gravitational, being rooted in the physics of inelastic materials.

The character of tidal dissipation is best seen by neglecting numerical multipliers and geometrical terms associated with the shapes of the orbiting bodies.  The shapes of most solar system bodies are unknown, so this neglect is at least partly reasonable. We consider for simplicity a binary object consisting of a primary of mass $m_p$ and a secondary of mass $m_s$, separated by a distance $r$, with $m_p \gg m_s$.  The radii of the primary and secondary objects are $a_p$ and $a_s$, respectively.    We also assume that the densities, $\rho$, of the primary and secondary are the same so that, neglecting shape specific factors, $m_p \sim \rho a_p^3$ and $m_s \sim \rho a_s^3$.

The gravitational pull of each body elongates the other along an axis connecting their centers, with the secondary being most deformed because of the greater mass and gravity of the primary.   If the deformation were instantaneous, this tidal bulge would stay perfectly aligned along the line of centers.  However, the primary in general rotates with an angular frequency, $\omega_p$, that is different from the orbital frequency of the satellite, $\omega_s$, and the tidal response is not instantaneous.  Empirically, most satellites orbit beyond the corotation radius, at which the orbital period is equal to the rotation period of the primary body. With $\omega_p > \omega_s$  the tidal bulge of the primary is carried ahead of the line of centers and the resulting torque on the primary acts to slow its rotation while expanding the orbit of the secondary.  In the opposite case ($\omega_p < \omega_s$) the tidal bulge lags behind, and the tidal torque increases $\omega_p$ while contracting the orbit of the secondary.  The latter is the case for Mars' satellite Phobos, which orbits at 2.76 R$_M$, far inside the corotation radius at 6.03 R$_M$ (1 R$_M = 3.4\times10^6$ m).  The orbit of Phobos will collapse into the planet because of tidal dissipation on a (surpringly short) timescale of a few 10s of Myr \citep{Bla15}.  Mars' other known satellite, Deimos, orbits beyond the corotation radius at 6.92 R$_M$ and is being slowly pushed away from the planet by tidal friction.  The finite response time and inelasticity of the material are central to the mechanism of tidal evolution because the misalignment of the tidal bulge creates an asymmetry upon which gravity can exert a torque.  

The gravitational force experienced from distance $r$ is $F = G m_p m_s/r^2$ or, equivalently, $F \sim G \rho^2 a_p^3 a_s^3/r^2$.  The gravity of the satellite  is slightly different on the near and far sides of the primary, by an amount

\begin{equation}
\delta F \sim \frac{dF}{dr} a_p \sim \frac{G \rho^2 a_p^4 a_s^3}{r^3}.
\label{deltaf}
\end{equation}

This small differential force periodically stretches and relaxes the bodies as they rotate and orbit, doing work in the process.   The average stress due to gravitational deformation is $S = \delta F/a_p^2$ [N m$^{-2}$].  The relation between the  stress and the strain (strain is the fractional change in the length scale, $s = \delta a_p/a_p$) is called the  bulk modulus, defined as $\mu = S/s$.  Substituting, the deformation is

\begin{equation}
\delta a_p \sim \frac{\delta F}{\mu a_p} = \frac{G \rho^2 a_p^3 a_s^3}{\mu r^3}.
\label{deltap}
\end{equation}

The work done by a force $\delta F$ applied over a distance $\delta a_p$ is $\delta W \sim \delta F \delta a_p$.  Substituting from Equations \ref{deltaf} and \ref{deltap} gives

\begin{equation}
\delta W = \left(\frac{G \rho^2 a_p^3 a_s^3 }{r^3}\right)^2  \frac{a_p}{\mu}
\label{loss}
\end{equation}

If the material were perfectly elastic, energy added to the body by stretching would be returned upon relaxation back to the original shape.  But in real materials, owing to internal friction, a fraction of the energy is dissipated as heat.  The fraction lost per stretching cycle is  conventionally defined as $Q^{-1} = \delta E/E$ (i.e., $Q$ is the inverse of what one would expect) and referred to as the tidal ``quality factor''.  High $Q$ corresponds to low dissipation per cycle and vice versa.  Given this, and recognizing that $\delta W$ in Equation \ref{loss} is the work done per cycle (not per second), we write the tidal power as

\begin{equation}
\frac{\delta W}{\delta t} \sim \left(\frac{G \rho^2 a_p^3 a_s^3}{r^3}\right)^2  \left(\frac{\omega_p}{Q}\right) \frac{a_p}{\mu}.
\label{loss2}
\end{equation}

\noindent The key point in all this is that the tidal power (Equation \ref{loss2}) scales with $r^{-6}$ because the differential tidal force, $\delta F$, and the deformation it causes, $\delta a_p$, are both proportional to $r^{-3}$.  

Finally, the timescale for dissipation to substantially change the rotational energy is $\tau_t \sim W /(\delta W/\delta t)$, where $W$ is the rotational energy.  We take  the rotational energy of the primary to be $W \sim I_p \omega_p^2$ where  $I_p \sim m_p a_p^2 \sim \rho a_p^5$ is the moment of inertia.  Then, 

\begin{equation}
\tau_t \sim \frac{ \mu Q \omega_p}{G^2 \rho^3 a_p^2} \left(\frac{m_p}{m_s}\right)^2 \left(\frac{r}{a_p}\right)^6
\label{tau_t}
\end{equation}

\noindent  is the order of magnitude time needed for torques from the satellite to substantially change the rotation of the primary.  Tides raised on the satellite by the gravity of the primary are larger, and act to change the rotation on an even shorter timescale, as can be seen by swapping $a_s$ and $a_p$ in Equation \ref{tau_t}.  Obviously, this derivation is highly simplified, but it serves to show the functional dependence of the tidal evolution timescale on material ($\rho$, $\mu$, $Q$) and geometrical ($a_p$, $a_s$, $r$) properties. 

\subsection{Internal Dissipation}
Even without an external force from a binary companion, internal energy dissipation in an inelastic material can modify the spins of individual asteroids.  The loss of rotational energy occurs when a body is rotating in a non-principal axis state, such that its rotational energy is not a minimum for its shape.  Then, any element of the body is subject to a cyclically varying stress as it rotates, which induces a variable strain, allowing internal friction to act.  The visual model is a deflected gyroscope, which both rotates around its axis and precesses and nutates at the same time.  

We can obtain a useful expression for the functional dependence of the damping timescale by the same dimensional method as used for the tidal effects in section \ref{tides}.  Precession and nutation of the rotation axis induce time-variable stresses that do work by cyclically compressing and relaxing the material.  The internal stress, proportional to the energy density in the body, is $S \sim \rho a^2 \omega^2$ [N m$^{-2}$], where $a$ is the nominal size of the body and $\omega$ [s$^{-1}$] the relevant angular frequency.  Stress and strain ($\delta a/a$) are related through the modulus $\mu = S/(\delta a/a)$, giving $\delta a \sim S a/\mu$. Then, by analogy with section \ref{tides}, the power dissipated as heat by the stress is $\delta W/\delta t \sim S a^2 \delta a (\omega/Q)$, which becomes

\begin{equation}
\frac{\delta W}{\delta t} = \frac{\rho^2 a^7 \omega^5}{\mu Q}.
\end{equation}

\noindent This compares with the instantaneous rotational energy $W \sim \rho a^5 \omega^2$, and the ratio of $W$ to $\delta W/\delta t$ gives the damping timescale

\begin{equation}
\tau_{damp} \sim A \frac{\mu Q}{\rho a^2 \omega^3}
\label{damp}
\end{equation}

\noindent where $A$ is a numerical multiplier introduced to account for aspects of material physics and asteroid shape that are not treated here.  \cite{Bur73} derived $A \sim$ 100, but values spanning a range of nearly three orders of magnitude have been reported in the literature, from $A$ = 1 to 4  \citep{Efr00}, to $A$ = 200 to 800 \citep{Sha05}, depending on details of the assumed visco-elastic properties and asteroid body shape.  Given the huge uncertainties in $A$, as well as those in $\mu$ and $Q$, the timescale offered by Equation \ref{damp} should be understood as no more than a qualitative guide, but the equation at least serves to show that damping by internal friction should be strongest in large bodies rotating rapidly.  This is consistent with the observation that a majority of the known non-principal axis (i.e., undamped) rotators are kilometer sized or smaller asteroids \citep{Pra05}.  

Assuming $\mu = 10^{11}$ N m$^{-2}$, $Q$ = 100, $\rho$ = 1000 kg m$^{-3}$ and a geometric middle value $A = 30$, we rewrite Equation \ref{damp} for the damping time in Myr as

\begin{equation}
\tau_{damp} \sim 10 \frac{P_r^3}{a^2}
\label{damp2}
\end{equation}

\noindent with $a$ in km and rotation period $P_r$ in hours.   A 1 km body with an excited rotation at $P_r$ = 3 hours would damp on timescale $\tau_{damp} \sim$ 300 Myr, by this relation.  The 0.1 km scale interstellar object 1I/'Oumuamua, with period $\sim$8 hour, has $\tau_{damp}$ longer than the age of the universe, consistent with photometric evidence that it might be in an excited rotational state \citep{Dra18}.   Larger asteroids whose rotation should be damped according to Equation \ref{damp2} can nevertheless be excited by impact, radiation torques or gravitational torques from close approaches to planets.  The nuclei of some comets also exhibit excited rotation, notably the 15 km long nucleus of 1P/Halley \citep{Sam91}.  In the comets, rotational excitation is a natural product of strong sublimation torques.  Again, we emphasize that the material properties $\mu$, $Q$ and $A$  (Equation \ref{damp}) are extremely poorly known (c.f., section \ref{matprop}, Figure \ref{muQ}), and damping times  very different from those given by Equation \ref{damp2} are possible.

\subsection{Examples}
\label{matprop}
\begin{figure}
\epsscale{0.99}
\plotone{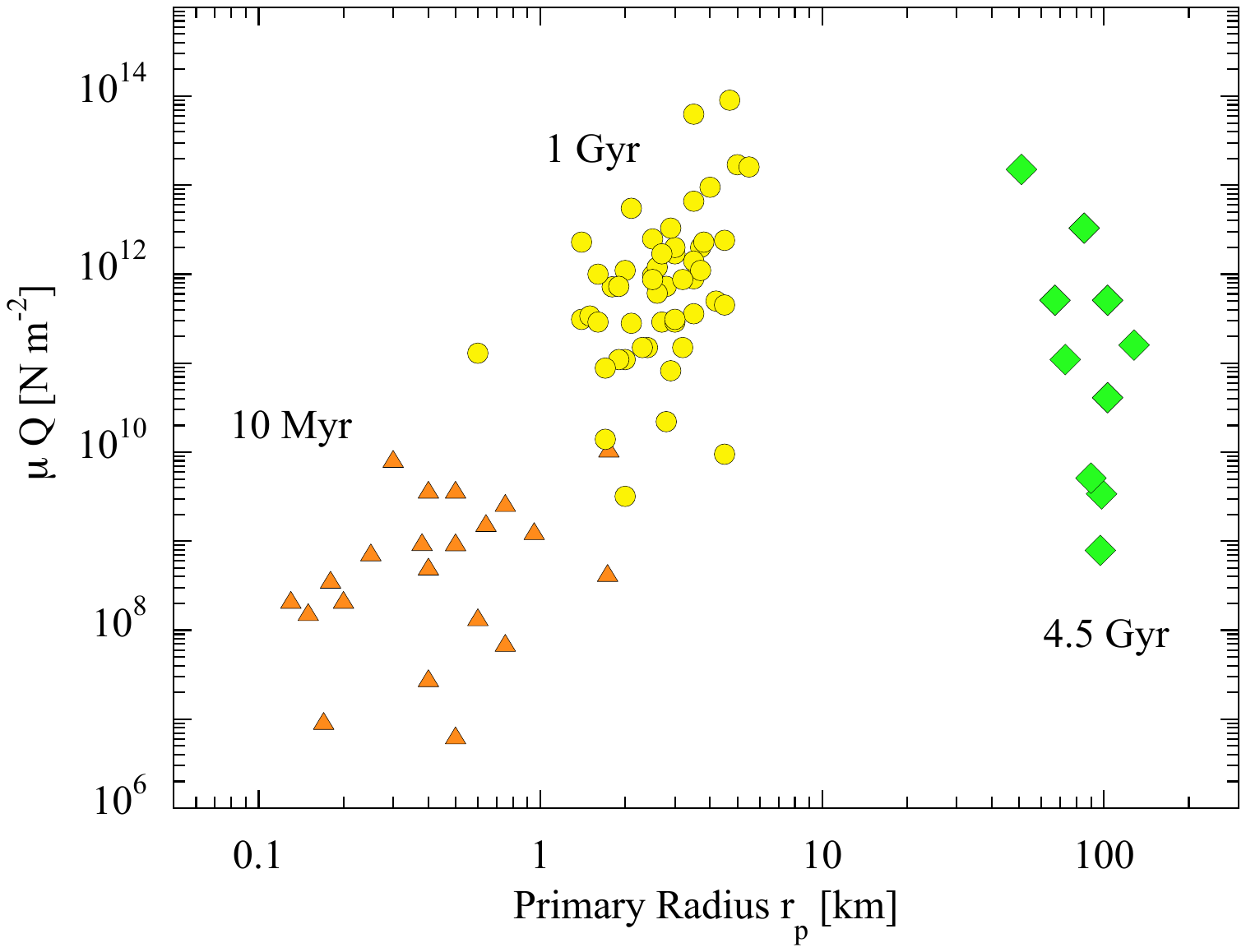}
\caption{$\mu Q$ as a function of radius estimated from binary asteroids.  The sub-kilometer objects (orange triangles) are near Earth asteroid binaries, for which a median age 10$^7$ year is assumed.  The $\sim$3 km asteroids in the main belt (yellow circles) are assumed to have median collisional age 10$^9$ year while the 100 km objects (green diamonds) are assumed to be 4.5 Gyr, as old as the solar system. Note the enormous scatter in $\mu Q$. Data from \cite{Tay11}.  
 \label{muQ}}
\end{figure}

\begin{figure}
\epsscale{0.99}
\plotone{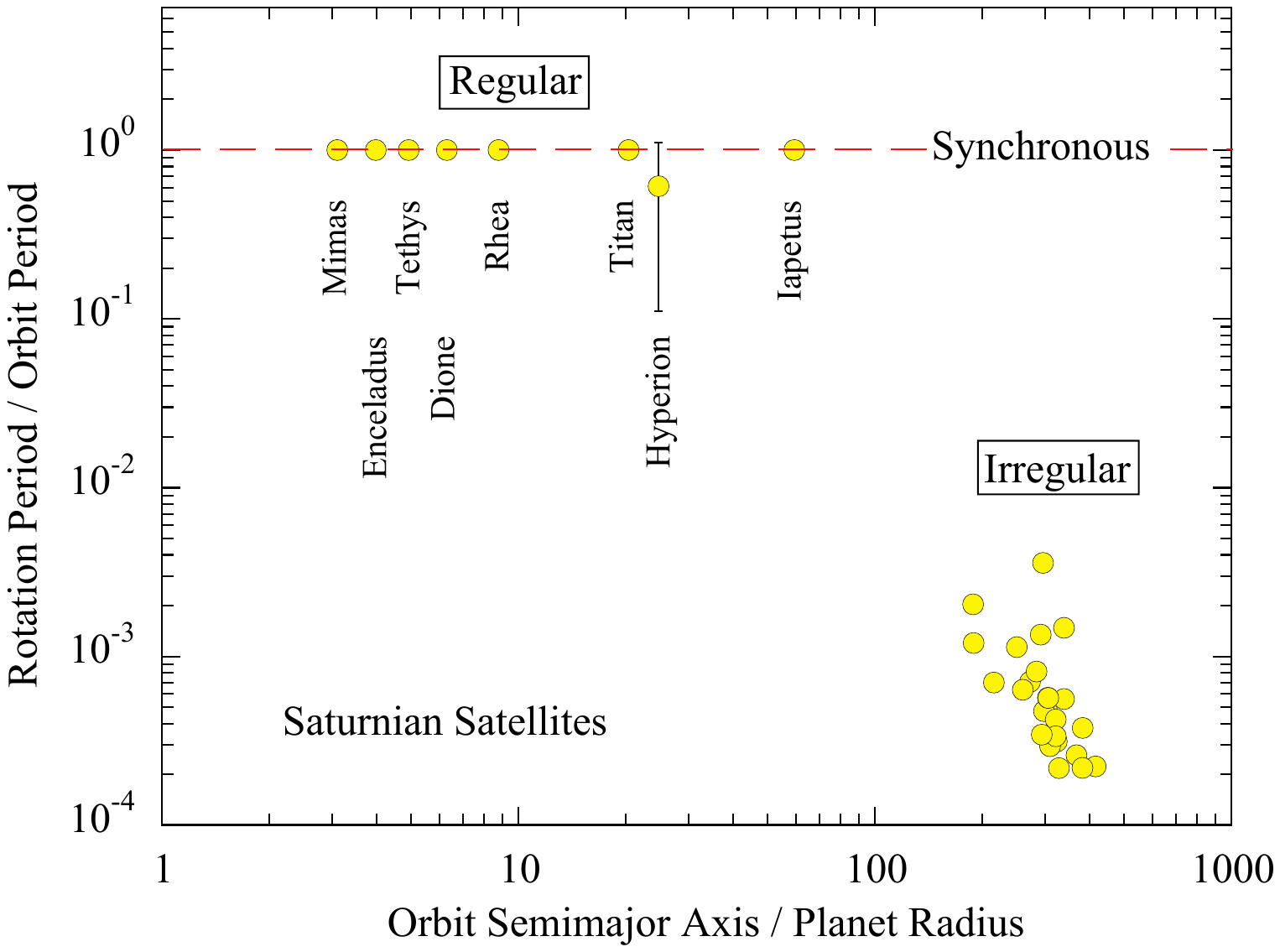}
\caption{Ratio of the rotation to orbit periods as a function of semimajor axis (in units of planet radius) for the measured satellites of Saturn.  The inner (regular) satellites rotate synchronously while the outer (irregular) satellites do not.  Hyperion is shown with an error bar to symbolically mark the fact that its rotation is chaotic.  
 \label{synchronous}}
\end{figure}

\begin{itemize}

\item \textbf{Material Properties:}   Neither $Q$ nor $\mu$ can be calculated from first principles for planetary bodies of interest.  Empirically, the quality factor varies widely (e.g., $Q \sim$ 10 for the Earth, $Q \sim$ 100 for the Moon \citep{Gol66}).  Values for gas giant planets are uncertain, but probably one or two orders of magnitude lower than $Q \sim 10^6$  initially estimated by these authors (c.f., \cite{Ful24}).  A value $Q \sim$ 100 is often assumed (albeit with little firm evidence) to apply to small solar system bodies.   Separately, the Young's moduli of terrestrial rocks span the range from $\mu \sim 10^9$ N m$^{-2}$ for sandstone and $\sim10^{11}$ N m$^{-2}$ for basalt to 4$\times10^{11}$ N m$^{-2}$ for diamond.  Many small bodies have an aggregate, ``rubble pile'' structure \citep{Wal18} that should give them mechanical properties quite different from solid rocks.  In particular, the porous and fragmented nature of asteroid rubble piles would seem to point by analogy towards $\mu$ values at the lower end of this terrestrial rock range, in which case $\mu Q \sim 10^{11}$ N m$^{-2}$ might be expected.  This value remains little more than a guess, and evidence from asteroid binaries suggests that even smaller $\mu Q$ values may prevail, as we next describe.

Useful empirical constraints on the product $\mu Q$ can be obtained from the application of Equation \ref{tau_t} to measurements of binary asteroids.  For example, if the age of a binary asteroid is known and the other physical properties (separation, period, component sizes, and density)  in the equation are measured or can be estimated, then substitution  gives $\mu Q$.  Figure \ref{muQ} shows $\mu Q$ obtained in this way as a function of primary asteroid radius for binaries in three size groups \citep{Tay11}. The median value for sub-kilometer near-Earth binaries (assuming $\tau_t = 10^7$ yr, the median dynamical lifetime of this population) is  $\mu Q = 5\times10^{8}$ N m$^{-2}$.   This is far smaller even than $\mu Q \sim 10^{11}$ N m$^{-2}$ estimated above. The median value for 3 km scale main-belt binaries (assuming $\tau_t = 10^9$ yr, the approximate collisional lifetime of objects this size) is $\mu Q = 7\times10^{11}$ N m$^{-2}$. Finally, $\mu Q = 1.4\times10^{11}$ N m$^{-2}$ for 100 km scale asteroids, computed assuming that they have survived for the 4.5 Gyr age of the solar system.  The considerable differences between these estimates may, in part, reflect a real size dependence of $\mu Q$ (c.f.~\cite{Nim19}), but they also testify to an element of guesswork in assigning values to some of the parameters in Equation \ref{tau_t}.  For example, it is difficult to assign more than a statistical age (e.g., based on the assumed collisional or dynamical lifetime) to any given binary object.

\item \textbf{Rotational Dissipation:} 
By Equation \ref{tau_t}, the tidal evolution timescale grows particularly strongly (as $r^{6}$) with the separation.  In a given system, doubling the separation increases the tidal timescale by a factor 2$^6$ = 64, almost two orders of magnitude.  This strong distance dependence is evident in real solar system objects.  For example,  a majority of satellites close to their parent planets  rotate synchronously, while the more distant  satellites do not.  Figure \ref{synchronous} plots the available measurements of the satellites of Saturn, showing the trend for close-in satellites to be rotationally synchronized while the rotations of distant satellites are unrelated to their orbital periods.  The 360 km diameter satellite Hyperion stands out.  Although Hyperion should be synchronized according to Equation \ref{tau_t}, its rotation is instead chaotic (i.e., the period changes irregularly) owing to impulsive torques exerted on its irregular body shape over the course of its eccentric ($e$ = 0.12) orbit, as predicted by \cite{Wis84} and established by \cite{Kla89a} and \cite{Kla89b}. 

The rotation of the Earth's Moon is obviously synchronized with its orbital motion and measurements show that the Earth's rotation is currently slowing (at $\sim10^{-5}$ s year$^{-1}$, or $\sim$3 hours per billion years \citep{Mit23}) as a result of tidal friction.  Inserting $r/a_p$ = 60, $m_p/m_s$ = 81, $\mu Q = 10^{11}$ N m$^{-2}$, $\rho = 5\times 10^3$ kg m$^{-3}$ and $a_p$ = 6400 km into Equation \ref{tau_t} gives $\tau_t \sim$ 0.5 Gyr.  Although this is only accurate to an order of magnitude, it is clear that tidal interaction with the Moon is responsible for slowing the spin of the Earth.  Likewise, terrestrial tides on the Moon slowed its rotation on a much shorter ($\sim$10 Myr) timescale, explaining its current synchronous rotation state.

\end{itemize}

\section{Yarkovsky Force}

\begin{figure}
\epsscale{0.99}
\plotone{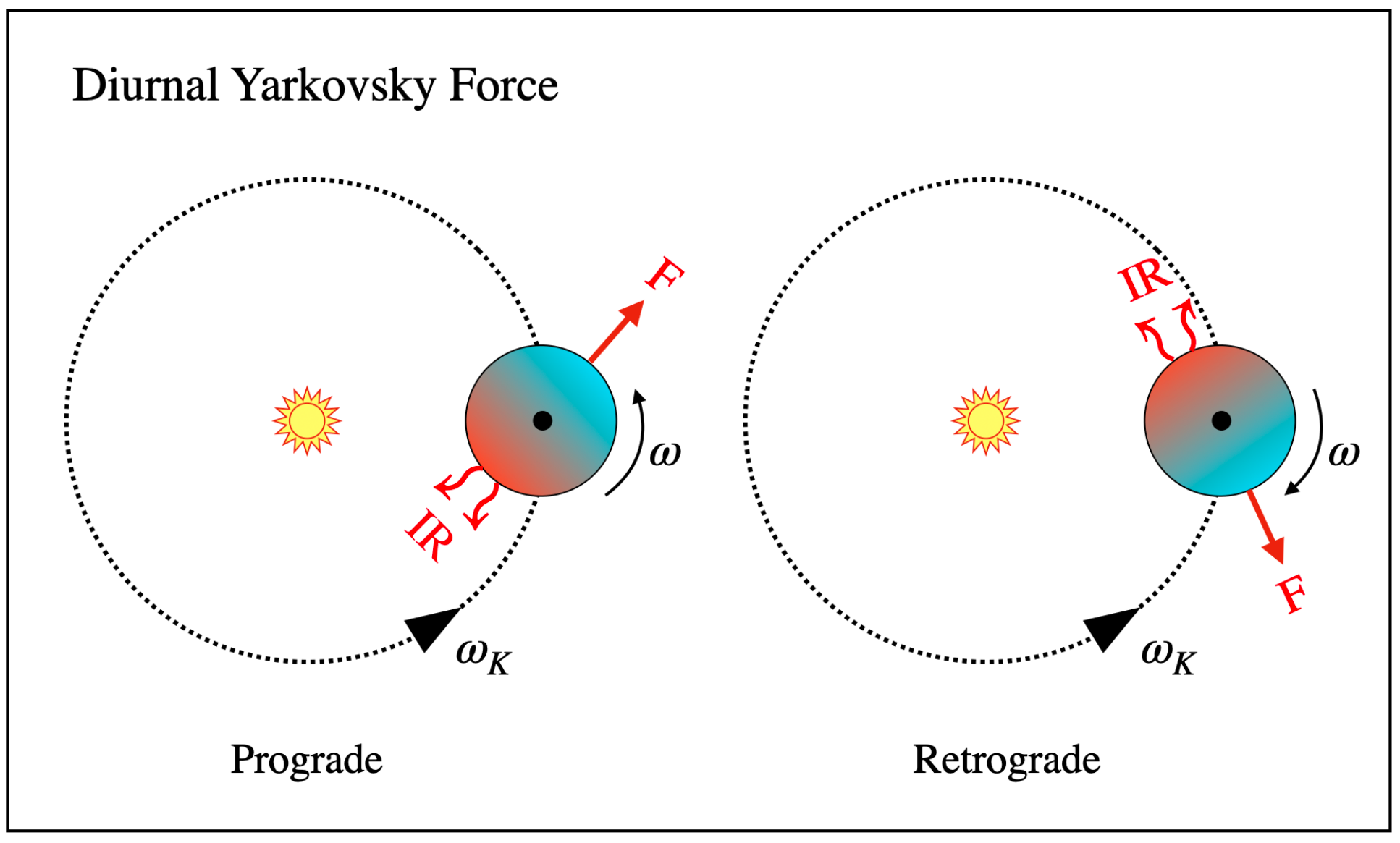}
\caption{Schematic plan view of a body orbiting the Sun, to illustrate the diurnal Yarkovsky force.  The dotted line shows the orbit of the body having angular rate, $\omega_K$, while the body itself rotates about an axis  (black dot) fixed perpendicular to the plane, with angular rate, $\omega$.    Rotation carries midday heat from the subsolar point into the afternoon where its loss by radiation is marked $IR$.  In the prograde case (left panel) $\omega_K$ and $\omega$ are parallel and the recoil force, $F$, acts with the orbital motion, causing the orbit to expand.  In the retrograde case (right panel), $\omega_K$ and $\omega$ are anti-parallel. The recoil force opposes the motion and causes the orbit to shrink. 
 \label{diurnal_yark}}
\end{figure}

An isothermal sphere at a finite temperature would radiate photons isotropically, experiencing no net recoil force from the loss of photon momentum.  However, real solar system bodies are neither spherical nor isothermal, being hotter on the sun-facing (day) side than on the opposite (night) side.  The distribution of  slopes, defined by both large and small scale features (e.g., boulders) on the surface, imbues the asteroid with a chirality, so that reflected and radiated photons exert a shape and surface texture dependent torque.  Moreover, body rotation, combined with the finite thermal response time of the surface material, carries peak heat from local noon into the afternoon, giving a ``thermal lag'' (see Appendix A). As a result of this thermal lag, the afternoon temperatures will always be higher than the corresponding morning temperatures on a rotating body. The chirality and the azimuthal thermal lag asymmetry in the temperature lead to a net force, the so-called diurnal Yarkovsky force.  The component of the force parallel to the direction of orbital motion can do work on the orbit, causing it to shrink or expand depending on whether the sense of the rotation is retrograde (i.e., opposite to the sense of orbital motion) or prograde (in the same sense as the orbital motion). See Figure \ref{diurnal_yark}. 

The above describes the diurnal Yarkovsky effect, which results from axial rotation of the asteroid and is maximized at 0\degr~and 180\degr~obliquity (a geometry assumed, for simplicity, in this discussion and in Figure \ref{diurnal_yark}).  For example, given a porous regolith with diffusivity $\kappa = 10^{-9}$ to 10$^{-8}$ m$^2$ s$^{-1}$, the thermal skin depth (Equation \ref{tcool}) on an asteroid rotating with period $P$ = 5 hours is $\ell \sim (P \kappa)^{1/2} \sim$ 4 to 13 mm.  At non-zero obliquities, the Yarkovsky force is reduced by a factor equal to the cosine of the obliquity. 

\begin{figure}
\epsscale{0.79}
\plotone{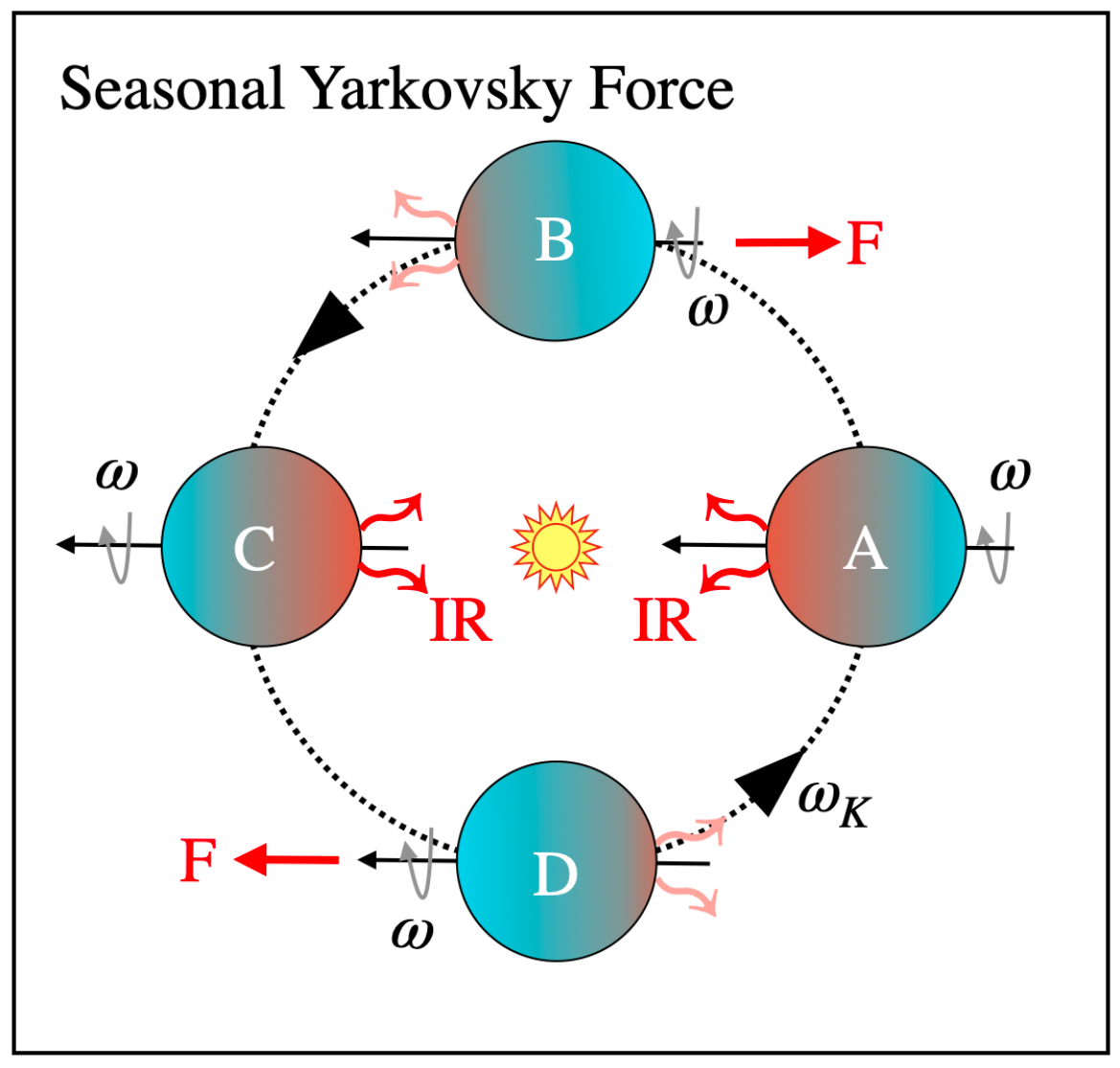}
\caption{Plan view to illustrate the seasonal Yarkovsky effect. The rotation vector (small black arrow, marked $\omega$), lies in the orbit plane and remains fixed in inertial space.  Peak insolation is reached when the spin vector points directly at the Sun, as at position A and shown by the strong curly red arrows. Heat is conducted into the interior and slowly leaks out as the asteroid moves around the orbit towards position B. The recoil force from the leakage of this residual heat (shown with faint curly red arrows), $F$, acts opposite to the motion.  Half an orbit later, the cycle repeats with the opposite hemisphere at position C, from which heat is retained to position D giving a recoil still opposing the orbital motion.  The seasonal Yarkovsky force always  opposes the orbital motion regardless of the sense of body rotation, resulting in orbital shrinkage
 \label{seasonal_yark}}
\end{figure}

The annual or seasonal Yarkovsky effect has a similar physical origin in thermal lag, but results from motion around the orbit and is maximized instead at 90\degr~obliquity (Figure \ref{seasonal_yark}).  One important, systematic difference between the seasonal and diurnal effects is that the former always opposes the orbital motion, causing a shrinkage of the orbit, whereas diurnal Yarkovsky can expand or contract it depending on the rotation direction.  A second difference is that, whereas the diurnal Yarkovsky effect results from thermal lag on asteroid rotation periods (typically hours), the seasonal effect results from thermal lag on the orbital timescale (typically years). The ratio of seasonal to diurnal timescales is years/hours $\sim$10$^4$ and, since the thermal skin depth scales with the square root of the time (Equation \ref{tcool}), the seasonal Yarkovsky effect depends on the thermophysical properties of a surface skin $\sim$10$^2$ times deeper than the diurnal Yarkovsky.  Instead of the diurnal Yarkovsky effect being driven by temperature variations in the top $\sim$ 4 to 13 mm of regolith, the seasonal effect is driven by an upper layer $\sim$0.4 to 1.3 m thick, all else being equal.  The relative magnitudes of the diurnal and seasonal Yarkovsky effects depend on many quantities, including the object size, spin vector and the depth dependence of the thermophysical properties of the surface layers (c.f.~Appendix A).  The diurnal effect typically dominates for bodies with regoliths (i.e., low diffusivity surfaces); we ignore the seasonal effect from further discussion here for simplicity. We also ignore the reflected component of the torque, because the albedos of asteroids are low (C-type albedos are a few percent, while even S-type albedos are only $\sim$20\%) and reflected torques are secondary.

Even so, the application to most real asteroids is problematic, because the force depends on many unknown or poorly constrained quantities. These include the magnitude and direction of the spin vector,  the body shape, the surface roughness and the thermophysical parameters responsible for the thermal lag (c.f.~Equation \ref{diffusion}).   For all these reasons, the Yarkovsky force cannot in general be calculated for a given body.  Instead,  it can be inferred from careful measurements of the action of the force.  The Yarkovsky force has been reviewed by \cite{Bot06} and \cite{Vok15}.  Here we offer an order of magnitude derivation that captures the essence of the process and we follow with some examples of its application.

The temperature of the Sun is $\sim$6000 K and its blackbody spectrum is peaked in the optical near 0.5 $\mu$m, whereas the isothermal blackbody temperature at 1 au is roughly 300 K and the spectral peak lies in the infrared near 10 $\mu$m.  The infrared photons carry 20 times less energy and 20 times less momentum than the absorbed solar photons, but they are 20 times more numerous so as to maintain energy balance on the body.  Therefore, we can use the radiation pressure exerted  by optical photons ($L_{\odot}/(4\pi r_H^2 c$)) to estimate the Yarkovsky force  due to thermal emission.  Specifically, we write the magnitude of the diurnal Yarkovsky force as

\begin{equation}
\mathcal{F}_Y = \frac{k_Y  \pi a^2}{ c} \left(\frac{L_{\odot}}{4\pi r_H^2} \right)
\end{equation}

\noindent where $0 \le k_Y \le 1$ is a dimensionless coefficient (to be determined) representing the fraction of the radiation pressure force acting parallel to the orbital motion.  (From the discussion in the preceding paragraphs, it should be evident that $k_Y$ is a function of the spin vector and thermophysical parameters and we are deliberately separating these from consideration here to simplify the presentation. Thermal considerations are briefly described in the Appendix).  Dividing by the mass of the body, assumed spherical, the resulting acceleration is

\begin{equation}
\alpha_Y = \frac{3 k_Y L_{\odot}}{16 \pi \rho c a r_H^2}.
\label{alphaY}
\end{equation}

\noindent We obtain an estimate for the Yarkovsky timescale from $\tau_Y \sim V_K/\alpha_Y$, $V_K$ being the Kepler speed.  Again assuming a circular orbit, we find

\begin{equation}
\tau_Y \sim \frac{16\pi \rho a c}{3 k_Y L_{\odot}} (G M_{\odot})^{1/2}  r_H^{3/2}.
\label{tau_yark}
\end{equation}

\noindent  The  order of magnitude radial drift can  be estimated from $d(r_H)/dt \sim r_H/\tau_Y$, or


\begin{equation}
\frac{dr_H}{dt} \sim \frac{3 k_Y L_{\odot}}{16\pi \rho a c } (G M_{\odot} r_H)^{-1/2},
\label{drift}
\end{equation}

\noindent  measured in m s$^{-1}$, and $dr_H/dt$ can be positive or negative, depending on the sense of asteroid rotation (c.f.~Figure \ref{diurnal_yark}).  The maximum possible drift rate at 1 au is given by setting $k_Y$ = 1, $r_H$ = 1 au, with the other parameters as above.  For $a$ = 1 km, this gives $d (r_H)/dt \sim 0.5$ km year$^{-1}$ or 4$\times10^{-3}$ au Myr$^{-1}$.  (We will show below that a more typical value is $k_Y \sim$ 0.05, so that $d (r_H)/dt \sim 2\times10^{-4}$ au Myr$^{-1}$ is a better estimate for a 1 km body at 1 au).  The asteroid belt is dynamically structured (e.g., at resonance locations) on scales $<$0.05 au, which can be crossed by kilometer-sized asteroids on timescales $<$250 Myr. We conclude that, depending on the size of the asteroid and the actual value of $k_Y$, the Yarkovsky force is capable of modifying the orbital properties  on timescales that are very short compared to the age of the solar system.  A practical limit to the influence of Yarkovsky drag is set by collisions with other bodies, which can  reset the spin and so change the magnitude and even direction of the radiative torque (\cite{Far98}, \cite{Wie15}).

\subsection{Examples}

\begin{itemize}

\item \textbf{Orbital Drift:} To obtain an estimate of the Yarkovsky constant, $k_Y$, we must use empirical data connecting measurements of the drift rate with the asteroid size, orbit and density. \cite{Gre20} and \cite{Fen24} report drift rate detections for 247 and 348 asteroids, respectively.  Figure \ref{rad_vs_drift} shows measurements of the magnitude of the radial drift rate as a function of radius for  58 near-Earth asteroids from \cite{Fen24}, having semimajor axes near 1 au and  formal signal-to-noise ratio $>$10.   Lines in the figure show Equation \ref{drift} with assumed values $k_Y$ = 0.01, 0.1 and 1. The data are reasonably well matched by the expected $1/a$ size dependence of the drift rate (Equation \ref{drift}) and by $k_Y \sim$ 0.02 to 0.13. We take $k_Y = 0.05$ as a representative value.   $k_Y \sim$ 0.05 implies a lag angle $\theta = \sin^{-1}(0.05) \sim$ 0.05 radians, or about 3\degr.   It should be remembered that the data are biased towards asteroids with the largest $k_Y$, because small values of $k_Y$  give  small and difficult to measure (low signal to noise ratio) drift rates.

\begin{figure}
\epsscale{0.99}
\plotone{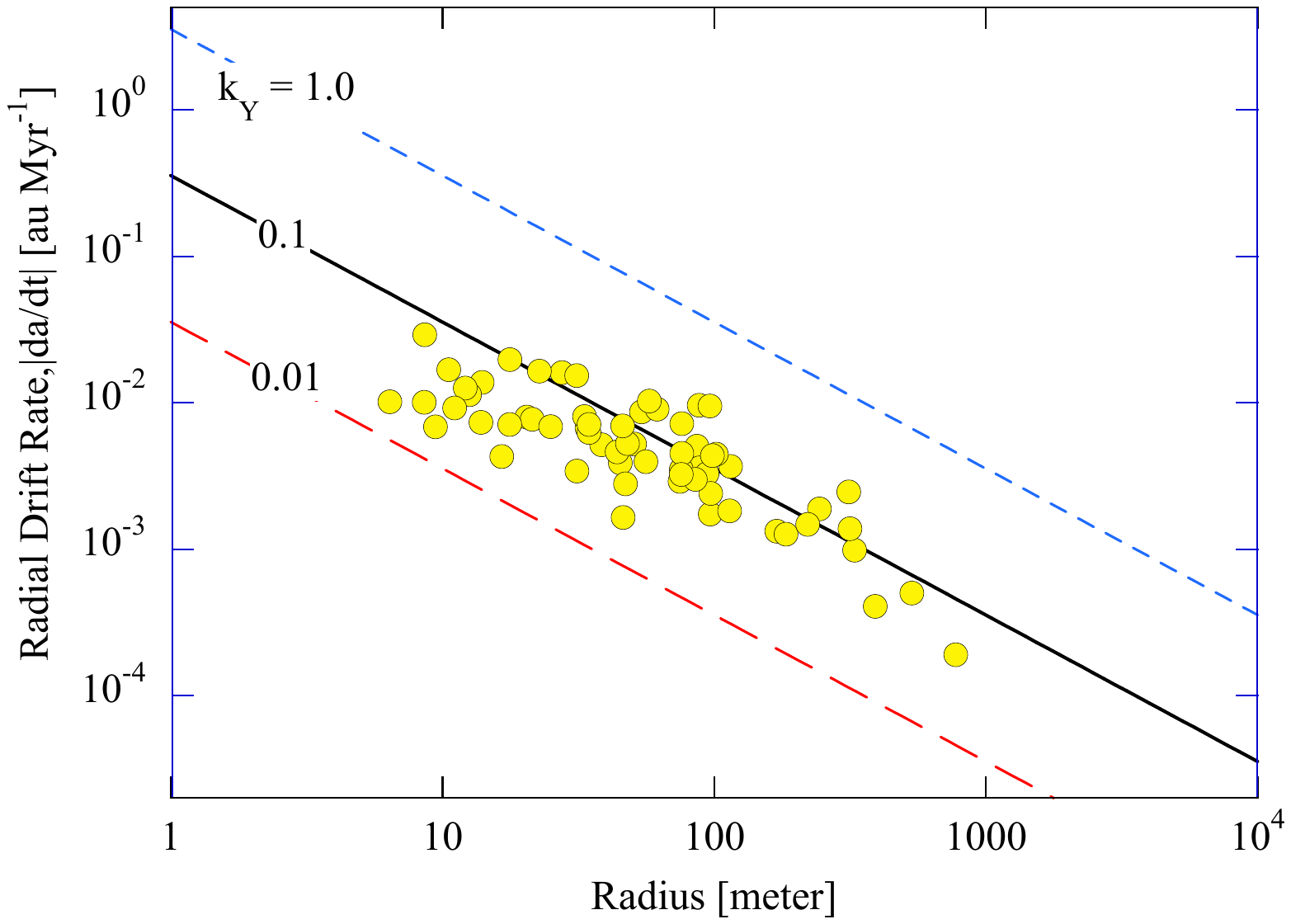}
\caption{Absolute value of the radial drift rate  plotted as a function of asteroid radius in meters. Lines show the model drift rate calculated from Equation \ref{drift} with $\rho = 10^3$ kg m$^{-3}$, $r_H$ = 1 au, $k_Y$ = 0.01 (red dashed line), 0.1 (solid black line) and 1 (dashed blue line).  Evidently, the data are broadly consistent with  $k_Y \sim$ 0.05.    Data are from the compilation by \cite{Fen24}.  
 \label{rad_vs_drift}}
\end{figure}

Given a random distribution of asteroid spin vectors, the above  model predicts that the ratio of positive (corresponding to prograde rotators) to negative (retrograde rotator) values of $dr_H/dt$ should be close to unity.  Instead, in both the studies by \cite{Gre20} and by \cite{Fen24}, fully 70\% of the high quality determinations have $dr_H/dt <$ 0, giving a ratio negative drift/positive drift = 2.3, and indicating an excess of  retrograde rotators. (Figure \ref{drift_rates}).  This inference matches measurements of the spins of near-Earth asteroids, which  show a preponderance of retrograde rotators (retrograde/prograde ratio $\sim$2)  \citep{Las04}. The bias in favor of retrograde rotation in the near-Earth population is thought to occur because  inward drifting main belt asteroids are more easily able to reach the $\nu_6$ secular resonance responsible for their deflection  into near-Earth space.

\begin{figure}
\epsscale{0.99}
\plotone{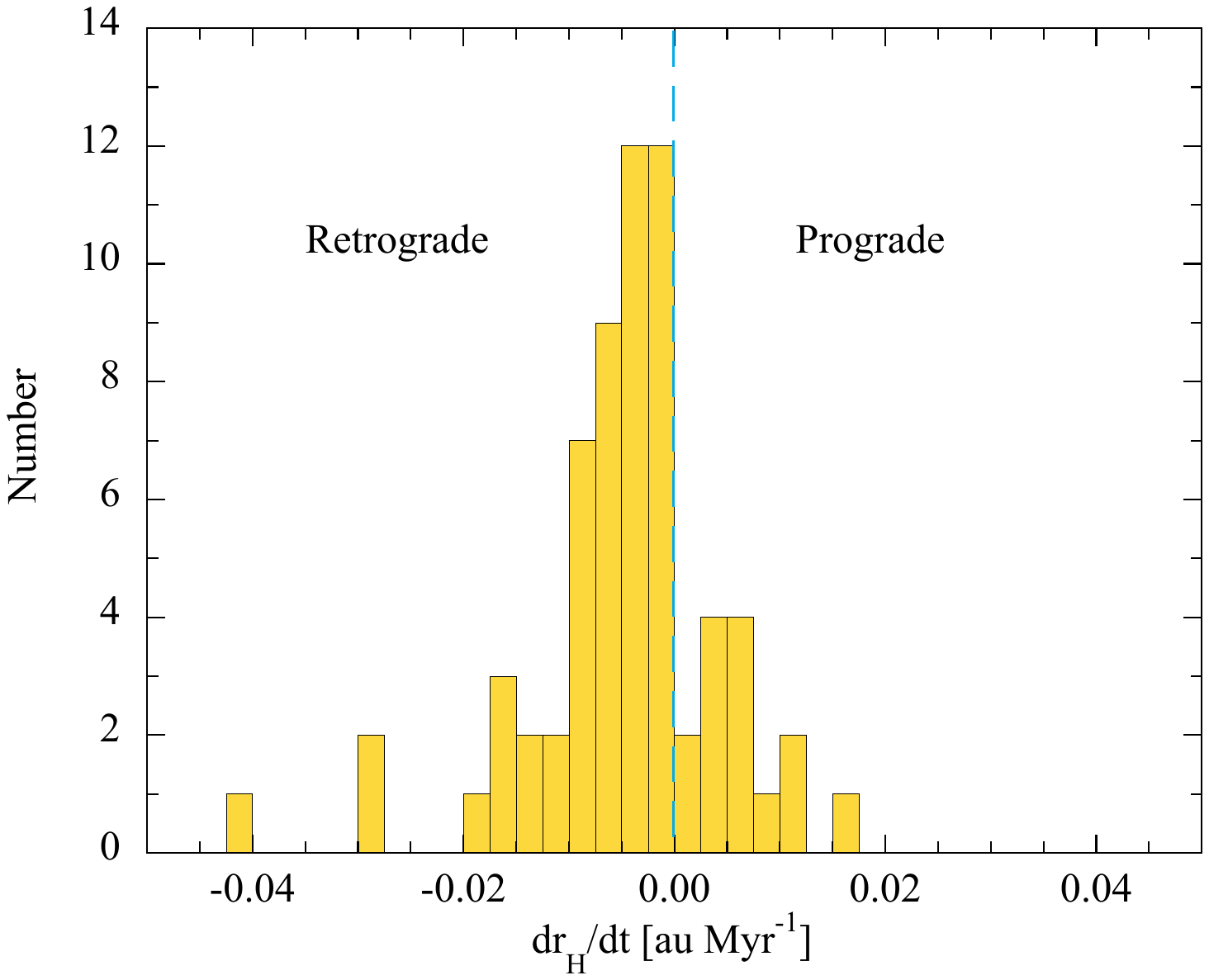}
\caption{Histogram of near-Earth asteroid radial drift rates showing an excess with negative values. The ratio of (inward drifting) retrograde rotators to (outward drifting) prograde rotators is $\sim$2.3:1. Data refer to sub-kilometer asteroids from \cite{Fen24}. Four objects with $dr_H/dt$ outside the plotted range have been excluded for clarity of the plot.  Asteroids with formal SNR $<$ 10 have not been considered.    
 \label{drift_rates}}
\end{figure}

\item \textbf{Origin of Meteorites:} Ultimately, Yarkovsky drift helps to supply the meteorites. The asteroid belt is crossed by numerous mean-motion and secular resonances, near which the orbits of bodies are  unstable.  Orbital eccentricities of resonant asteroids are excited until they become planet-crossing and short-lived.  The mean-motion resonances would be nearly empty if it were not for Yarkovsky drift, which is responsible for feeding resonance regions with nearby asteroids through semimajor axis drift \citep{Far98}.  About 80\% of the near-Earth objects are supplied from the $\nu_6$ secular resonance   and, spectrally, 2/3 of these are S-type asteroids \citep{San24}.  The S-types are related to the thermally metamorphosed LL chondrite meteorites, themselves fragments of bodies from the inner belt.

\begin{figure}
\epsscale{0.9}
\plotone{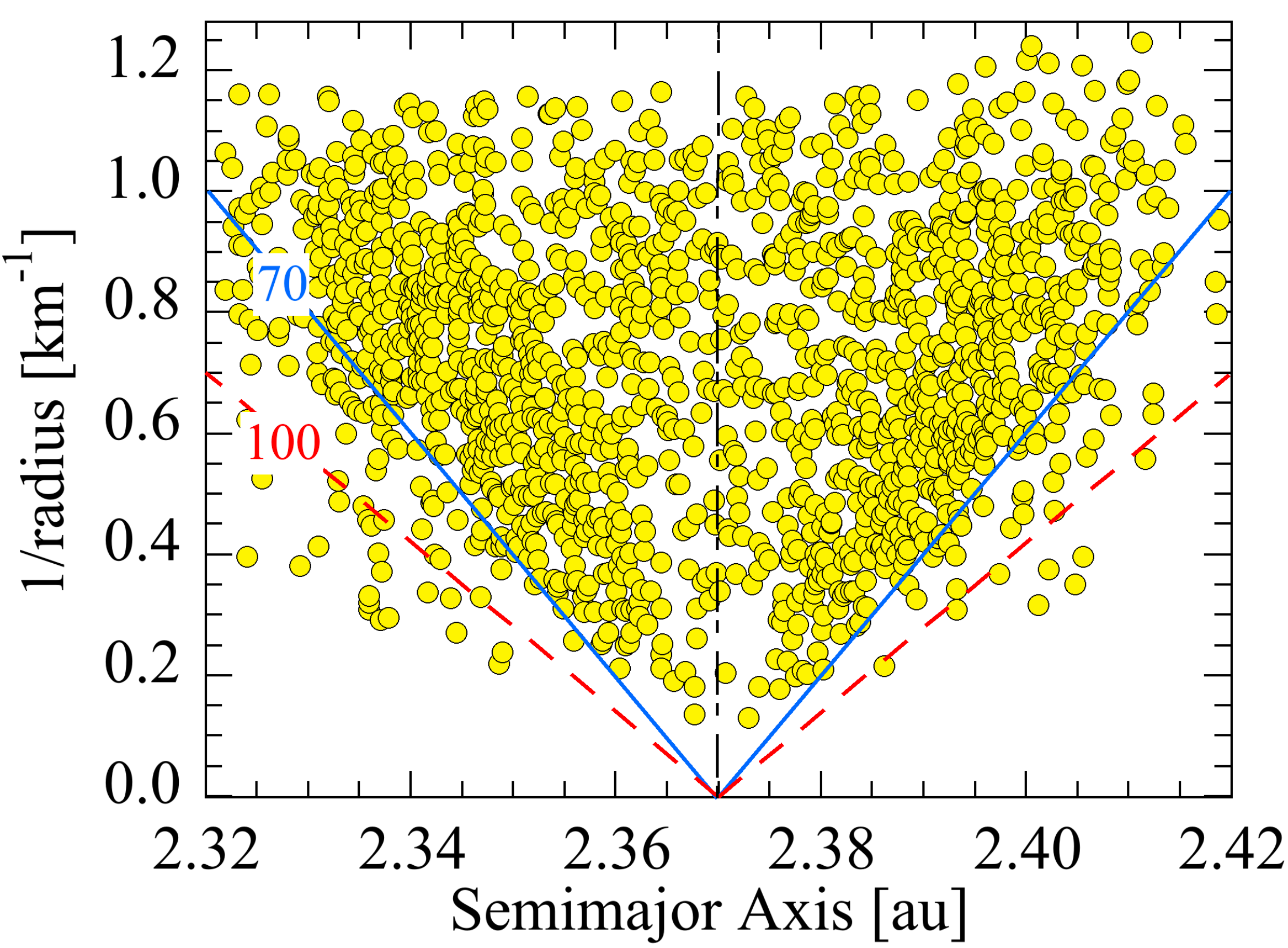}
\caption{Yarkovsky spreading diagram for the Erigone main belt asteroid family.  The vertical dashed black line shows $r_H$ = 2.37 au, the semimajor axis of the presumed family parent body.  Solid blue and red dashed  lines show trajectories for family ages 70 Myr and 100 Myr. Data from \cite{Bol18}, models from Equation \ref{drift}.      
 \label{bryce}}
\end{figure}

\item \textbf{Dispersal of Asteroid Families:}
Shattering collisions between asteroids produce families of objects having initially similar orbital elements.  However, family asteroids experience a size dependent drift from their initial orbital semimajor axes under the action of the Yarkovsky force.  By Equation \ref{drift}, the semimajor axis drift in time $\Delta t$ is $\Delta r_H \propto \Delta t/a$, and $\Delta r_H$ can be positive or negative depending on the sense of rotation of the asteroid.  Small asteroids drift the farthest, all else being equal, giving rise to a V-shaped distribution in a plot of semimajor axis vs.~$1/a$.  This effect is observed in the main asteroid belt, where it has been used to identify families and estimate their ages.  An example is shown in Figure \ref{bryce} which compares data for the Erigone family with Equation \ref{drift} (we assumed $k_Y$ = 0.05, $\rho$ = 2000 kg m$^{-3}$ and initial semimajor axis $r_H$ = 2.37 au). There is some ambiguity in defining the edges of the distribution, but spreading ages near 70 Myr to 100 Myr are plausible, while much larger and smaller ages are not. \cite{Bol18} estimated an age of 90 Myr. The method is not perfect. The initial orbits may not have exactly the same semimajor axis, the effect of  asteroid obliquity is not included (Equation \ref{drift} gives the maximum value, for obliquity = 0\degr) and the individual obliquities are in any case not known, the sizes and densities of asteroids are approximate, and external perturbations (e.g., from  planets) can act to spread the orbits (especially for the older clusters).   Nevertheless, Yarkovsky spreading gives an impressive explanation for the V-shaped diagram and a useful measure of the family age.  

\item \textbf{Post-Main Sequence Evolution of the Sun:} In $\lesssim$5 Gyr the Sun will exhaust its core hydrogen and bloat into a red giant, with a luminosity $\sim10^{3.5}$ times its present value \citep{Vas93}.  When this occurs, the Yarkovsky force will likewise jump by a factor $\sim$10$^{3.5}$ causing accelerated orbital drift to compete with many other processes to destabilize small orbiting bodies of the solar system.  Destabilized planetary systems are likely implicated in feeding debris to the metal polluted central white dwarf remaining after the red giant phase \citep{Ver19}.  

\item \textbf{{Binary Yarkovsky:}}
The Yarkovsky force can  affect a binary in two ways.  First, in a binary with small separation, thermal radiation from each component can heat the other, resulting in a kind of radiation drag distinct from the Yarkovsky drag caused by direct sunlight.  Second, if the  orbital inclination is also small, the two components will pass through each other's shadow in each orbit.  Since the thermal response of the surface material is lagged (see Appendix A), the entry into and exit from the shadow introduces an additional asymmetry into the radiative force that results in yet another torque.  These subtle effects, the former somewhat unfortunately named ``planetary Yarkovsky'' and the latter ``Yarkovsky-Schach'' drag,  may rival tidal dissipation (Section \ref{tides}) and BYORP (Section \ref{YORP2}) in the orbital evolution of some close binary asteroids \citep{Zho24}.  

\end{itemize}

\section{Lorentz Force}
\label{Lorentz}

A charged particle interacts with a moving magnetic field through the Lorentz force, $\mathcal{F}_L$. For simplicity, we ignore the fact that $\mathcal{F}_L$ is a vector acting perpendicular to both the velocity of the particle and the direction of the field, and we consider only the magnitude of the force, $\mathcal{F}_L = B q V$. Here $B$ is the magnetic flux density, $q$ is the charge on the particle and $V$ is the velocity of the particle with respect to the field. 

To evaluate the Lorenz force we first need to know the charge on a particle. Coulomb's Law gives the force between two  charges, $q$, separated by distance, $r$, as $\mathcal{F} = q^2/(4\pi\varepsilon_0 r^2)$, where $\epsilon_0 = 8.854\times10^{-12}$ F m$^{-1}$ is the permittivity of free space. The work done in bringing a charge $q$ from $r = \infty$ to the surface of a particle of radius $r = a$ is just $E = \int_{\infty}^a \mathcal{F} dr = q^2/(4\pi \varepsilon_0 a)$.  The Volt is a measure of the work done, $E$, in moving a charge, $q$, through a potential difference, $U$.  Then, $U = E/q$ gives the relation

\begin{equation}
q = 4\pi \epsilon_0 U a
\label{q}
\end{equation}

\noindent for the charge on a particle of radius $a$ when its potential, $U$, is known. Numerous effects contribute to the charging of dust particles in the solar system, but the dominant effect is photoionization from solar ultraviolet, which leads to $U \sim$  5 to 10 Volts \citep{Wya69} (corresponding to photon wavelengths $\sim$1000\AA~to 2000\AA). This potential can vary depending on fluctuations in the solar ultraviolet flux, itself a function of the $\sim$22 year solar magnetic cycle. The potential, $U$, depends on the photon energy and the ionization threshold of the material, and is therefore approximately independent of distance from the Sun.  However, the photon flux varies as $r_H^{-2}$ and so the charging time increases as $r_H^2$; it is $\sim 10^3$ times longer in the Kuiper belt at 30 au than near the Earth at 1 au, elevating the relative importance of leakage currents that ultimately can limit the accumulated charge.  In the following we take $U$ = 10 Volts independent of distance.

At distances $r_{au} \gg$ 1 the Sun's Parker spiral is wound so tightly as to be nearly azimuthal.  In the planetary region, the azimuthal magnetic flux density approximately follows $B(r_H) = B_1 r_H^{-1}$, with $B_1 \sim$ 600 T m and $r_H$  expressed in meters (as determined from Figure 11 of \cite{Bal13}), while the field is swept with the solar wind at speed $V \sim$ 500 km s$^{-1}$, approximately independent of heliocentric distance.  The field does exhibit substantial fluctuations through the solar cycle and also varies with heliographic latitude; we ignore these effects here for simplicity.  Substituting for $q$ from Equation \ref{q} and for $B$ and dividing by the particle mass gives the particle acceleration due to the Lorentz force as

\begin{equation}
\alpha_L = \frac{3 B_1 \varepsilon_0 U V}{\rho a^2 r_H}.
\end{equation}

\noindent Further dividing by the local gravitational acceleration (Equation \ref{gsun}) defines the magnetic $\beta$ parameter $\beta_L = \alpha_L/g_{\odot}$

\begin{equation}
\beta_L = \frac{3 B_1 \varepsilon_0 U V}{ G M_{\odot} \rho}\left(\frac{r_H}{a^2}\right).
\label{betaM}
\end{equation} 

\begin{figure}
\epsscale{0.99}
\plotone{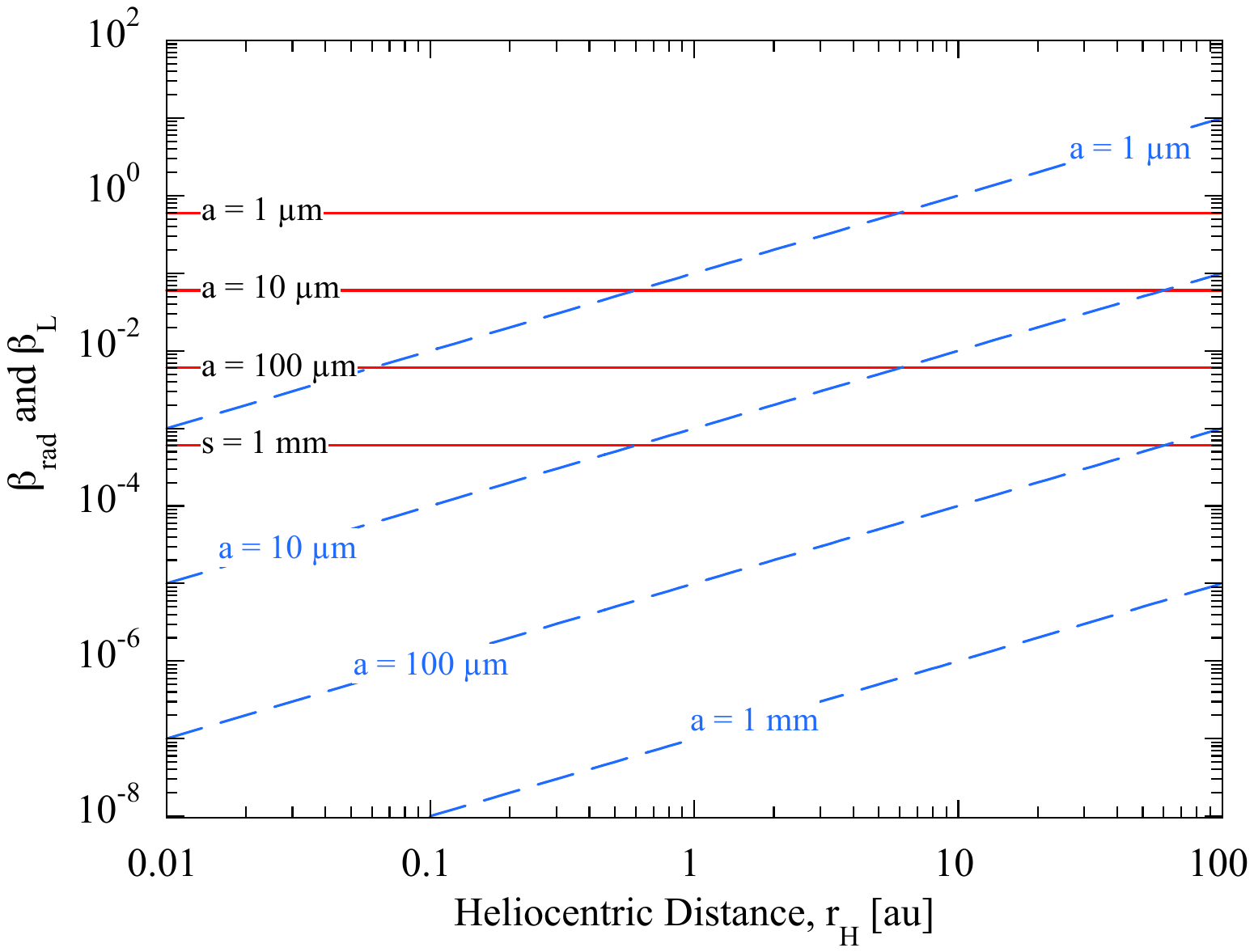}
\caption{$\beta_{rad}$ (solid red lines) and $\beta_L$ (dashed blue lines) for particles with radii 1, 10, 10$^2$ and 10$^3$ $\mu$m and density $\rho = 10^3$ kg m$^{-3}$, as functions of heliocentric distance.  $\beta_L > \beta_{rad}$ only for the smallest particles at the largest heliocentric distances.
 \label{beta_plot}}
\end{figure}

Substitution gives

\begin{equation}
\beta_L \sim 0.1 \left(\frac{r_{au}}{a_{\mu}^2}\right)
\label{betaM2}
\end{equation}

\noindent with $r_{au}$ being the heliocentric distance expressed in au and $a_{\mu}$ the particle radius expressed in $\mu$m.  Figure \ref{beta_plot} shows $\beta$ and $\beta_L$ as functions of heliocentric distance for particles with density $\rho = 10^3$ kg m$^{-3}$ and 1, 10, 10$^2$ and 10$^3$ $\mu$m in radius. Equations \ref{betaM} and \ref{betaM2} show that the dynamical importance of the Lorentz force is maximized for the smallest particles and the largest heliocentric distances.  

\subsection{Examples}

\begin{itemize}

\item \textbf{Interstellar Dust:} Spacecraft with velocity-measuring impact detectors record dust particles traveling faster than the local gravitational escape speed from the Sun \citep{Gru85}.  These are of interstellar origin but are, on average, substantially larger than the interstellar dust known from astronomical measurements of extinction and polarization. The impact energies correspond to particles several tenths of a micron (up to $\sim$0.4 $\mu$m) in size, whereas the extinction and polarization signatures of the interstellar medium are due to dust with sizes down to $a \sim$ 10 nm. Some particles are deflected by radiation pressure but, as $a \rightarrow$ 0, $Q_{pr} \ll$ 1, leaving  the Sun's magnetic field to deflect the smallest and most abundant \citep{Man21}.  For example, consider dust approaching the heliopause at 25 km s$^{-1}$, where $B \sim$ 0.5 nT \citep{Bur22}. These parameters give $\beta_L >$  1 for $a <$ 0.1 $\mu$m, so smaller particles are deflected from the planetary region by the heliospheric magnetic field, consistent with dust counter measurements \citep{Ste13}. The penetration distance of interstellar dust is a function not only of particle size but also of time, because of the Sun's 22 year magnetic cycle \citep{Lan00}.

\item \textbf{Comet Dust:} The  Lorentz force acts  roughly perpendicular to the Sun-comet line, allowing the possibility that the dust morphology of a comet might be altered by the Lorentz force relative to  that expected on the basis of radiation pressure and solar gravity alone.

The ratio of the Lorentz acceleration (Equation \ref{betaM2})  to the radiation pressure acceleration (Equation \ref{beta2}) in the geometric optics limit is

\begin{equation}
\frac{\beta_L}{\beta} \sim 0.2 \left(\frac{r_{au} }{a_{\mu} }\right).
\label{betaratio}
\end{equation}

\noindent

Near 1 au, $\beta_L/\beta \ll$1 for the micron sized and larger particles that dominate the scattering, providing justification for the neglect of Lorentz force in models of  comets in the terrestrial planet region.  Equation \ref{betaratio} suggests that the effects of Lorentz force should be more significant at larger heliocentric distances where, unfortunately, we possess  few relevant observations.  

Three exceptions are the long-period comets  C/2014 B1 (Schwartz),  C/2010 U3 (Boattini) and C/1995 O1 (Hale-Bopp).  C/2014 B1 (Schwartz) was dominated by  quite large ($a_{\mu} >$ 100) particles at $r_{au} \sim$ 10  \citep{Jew19}. It has $\beta_L/\beta \lesssim 0.02$ by Equation \ref{betaratio}, and showed no effect from Lorentz force.  Observations of  C/2010 U3 (Boattini) at larger distance ($r_{au} \sim$ 20 au)  showed smaller  particles ($a_{\mu}$ = 10) \citep{Hui19}.  Equation \ref{betaratio} gives $\beta_L/\beta \sim$ 0.4, consistent with a larger but still not dominant influence of the Lorentz force.  C/1995 O1 (Hale-Bopp) was observed at $r_{au} \sim$ 20 and reported to show a coma of small particles with $a_{\mu} \sim$ 1 \citep{Kra14}.  The coma could not be well matched by models using only radiation pressure.  With the above values, Equation \ref{betaratio} gives $\beta_L/\beta \sim$ 4, consistent with a Lorentz force dominated morphology.   In the Kuiper belt, at $r_H$ = 40 au, the motion of all particles smaller than about 8 $\mu$m, including the micron-sized particles detected by the New Horizons particle counter \citep{Don24}, should be affected by the Lorentz force.

At very small particle sizes, $a \ll \lambda$, the geometric optics limit implicit in Equation \ref{betaratio} breaks down, and the radiation force is reduced by a factor $Q_{pr} < 1$, increasing $\beta_L/\beta$. The motion of particles $a \lesssim$ 100 nm is strongly affected by Lorentz forces \citep{Man21}.

\item \textbf{Dust in Planetary Magnetospheres:}
The giant planets sustain strong dynamo-generated magnetic fields and provide numerous examples where magnetic forces are important. For example, the equatorial surface field of Jupiter is $B_J \sim$ 400 $\mu$T, about 10$^5$ times stronger than the solar wind field near Earth, and magnetic forces are of correspondingly greater significance. Strong magnetospheric fields are able to retain plasma that would otherwise be quickly stripped away by the solar wind, creating a local gas environment that is distinct from the open interplanetary medium.  The magnetospheres of  Jupiter and Saturn are filled with dense plasma ($\sim 10^8$ to 10$^9$ m$^{-3}$, or 10$^2$ to 10$^3$ times the solar wind density)  from the outgassing satellites Io and Enceladus, respectively.  Dust near these satellites is charged negatively by contact with the  plasma (because the electrons travel faster in thermal equilibrium and so deliver a larger charging current than the heavier, slower ions).  The charge reverses to positive values at larger distances where the effects of plasma charging are overcome by those of UV photoelectron loss \citep{Spa19}.  Charged dust particles are accelerated by the planetary field, which sweeps past at the corotation speed. One consequence of magnetospheric dust interactions is that Jupiter \citep{Gru93} and Saturn \citep{Kem05} eject streams of nanometer sized particles (mass $\sim10^{-24}$ kg, magnetic $\beta_L \gg$ 1) at speeds above the gravitational escape speed from either planet \citep{Hsu12}. Another is that the B ring of Saturn displays transient, 10$^4$ km scale, quasi-radial dust structures known as spokes, whose abundance is seasonally modulated by the ring illumination but also by Saturn's rotating magnetic field (\cite{Smi81}, \cite{Por82}).  The origin of spokes is still not fully understood, but it is clear that they consist of micron-sized dust particles briefly elevated above the Saturn ring plane and with motions that reflect the combined influence of local electrostatic and magnetic forces, as well as planetary gravity \citep{Hor04}.

\end{itemize}

\subsection{Summary}
\begin{figure}
\epsscale{0.6}
\plotone{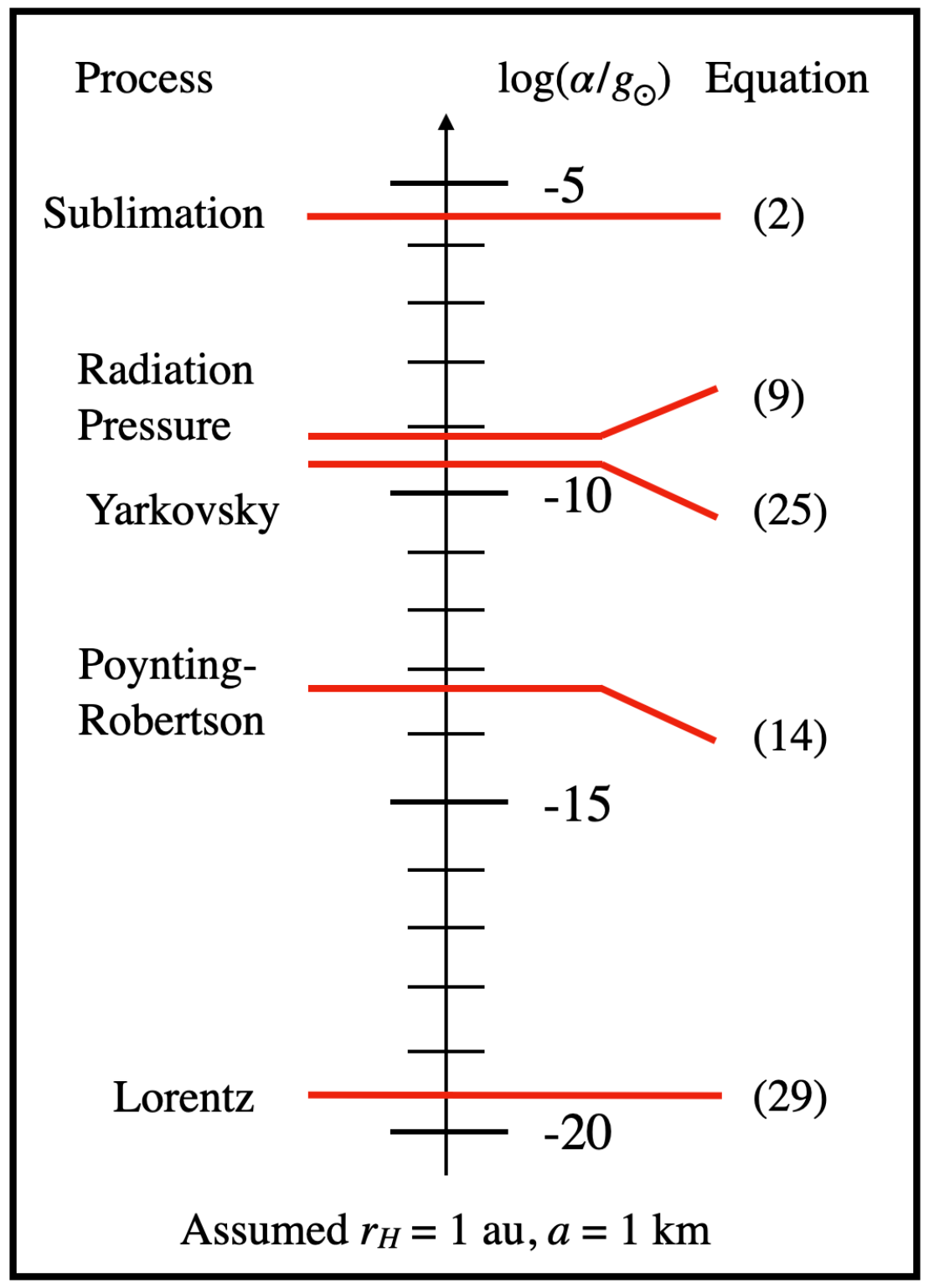}
\caption{Relative magnitudes of the non-gravitational accelerations discussed in the text, relative to solar gravity, for $r_H$ = 1 au and object radius $a$ = 1 km.
 \label{pressure}}
\end{figure}

The relative magnitudes of the accelerations are compared in Figure \ref{pressure}.  For this purpose, we evaluate each acceleration at $r_H$ = 1 au and for a body radius $a$ = 1 km, and we divide by the local solar gravitational acceleration ($g_{\odot}$ = 0.006 m s$^{-2}$ at $r_H$ = 1 au).    
 
\section{Torques}
Most asteroids and cometary nuclei are  irregular in shape. On such objects, the non-gravitational forces described above generally do not pass through the center of mass, resulting in a torque.  Left unchecked, the torque will drive inexorably towards rotational instability, where the centripetal forces exceed the gravitational and cohesive forces holding the body together.  A strengthless oblate ellipsoid of density $\rho$ and with equatorial axes $a$ = $b$ and polar axis $c \le a$ has critical period

\begin{equation}
\tau_{crit} = \left(\frac{3\pi}{G\rho}\right)^{1/2} \left( \frac{a}{c}\right)^{1/2}.
\label{dens}
\end{equation}

\noindent For a sphere ($a = c$) of density $\rho = 10^3$ kg m$^{-3}$, Equation \ref{dens} gives  $\tau_{crit}$ = 3.3 hour.  A highly oblate body of the same density but with $a/c$ = 2 would have a critical period 2$^{1/2}$ times longer, or 4.7 hours.   A majority of the measured rotation periods of solar system bodies are within a factor of a few of $\tau_{crit}$ \citep{War09}, suggesting the importance of centripetal forces. However, it must be admitted that observational biases against the measurement of periods either much shorter or much longer than a few hours are strong in most existing datasets.

In general, applied torques change both the direction and the magnitude of the spin vector, inducing excited or non-principal axis rotation.  Observationally, precessional effects tend to be subtle, requiring longer sequences of observation than are commonly available.  Therefore, in the following discussion we concentrate on the more easily measured changes to the magnitude of the spin (i.e., to the rotational period) for which we already possess abundant evidence. Torques due to sublimation and to radiation are of particular significance for the destruction of comets and asteroids, respectively. We consider them separately next.

In both cases a simple dimensional treatment is informative.  We note that the timescale for an applied torque, $T$, to change the spin is just $\tau_s \sim L/T$, where $L$ is the angular momentum of the body.  For a homogeneous sphere of radius $a$ and density $\rho$, the angular momentum is $L  = 16\pi^2 \rho a^5/(15 P)$, where $P$ is the instantaneous rotation period.

\subsection{Sublimation Torque}
The magnitude of the sublimation torque on a comet is equal to the momentum lost per second in sublimated material multiplied by the moment arm, which we write as $k_T a$, where $a$ is the nucleus radius and 0 $\le k_T \le 1$ is a dimensionless multiplier.  $k_T$ = 0 corresponds to perfectly isotropic sublimation with no net torque.  $k_T$ = 1 corresponds to perfectly collimated ejection tangential to the surface of the body.  Then, with a mass loss rate $\mu m_H Q_g$ [kg s$^{-1}$] and an outflow speed $V_{th}$, the magnitude of the torque is $T = \mu m_H Q_g k_T V_{th} a$. The timescale for the torque to change the spin is then \citep{Jew21b}

\begin{equation}
\tau_s \sim \frac{16\pi^2 \rho a^4}{15\mu m_H Q_g k_T V_{th}} \frac{1}{P}.
\label{tau_s}
\end{equation}

\noindent Note that, for a fixed $Q_g$, Equation \ref{tau_s} has a very strong ($a^4$) nucleus radius dependence.  However, it is  natural to expect that the production rate, $Q_g$, should scale with the nucleus area, $Q_g \propto a^2$ and, if it does, $\tau_s \propto a^2$ is anticipated for the nuclei of comets, all else being equal.  

\begin{figure}
\epsscale{0.9}
\plotone{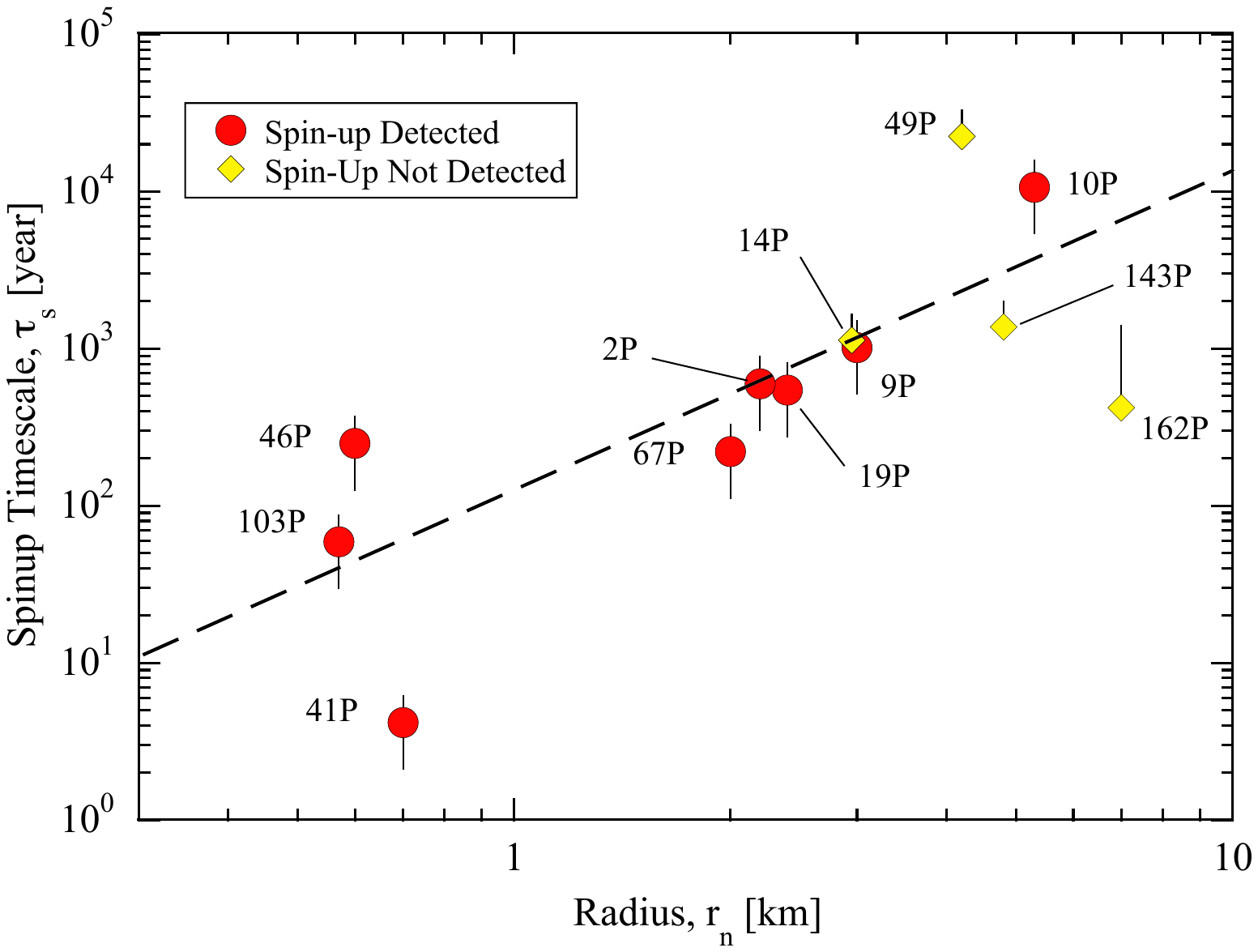}
\caption{Measured timescale for changing the rotation period, $\tau_s$, as a function of the nucleus radius for short-period comets with perihelia in the 1 to 2 au range.  Filled red circles show spin change detections while yellow diamonds show lower limits to the allowed spinup timescales. Comets are identified by their numerical labels.  The dashed line shows $\tau_s = 100 a^2$. From \cite{Jew21b}. }
\label{comets_spin}
\end{figure}

The rotational lightcurves of some comets have enabled the measurement of spin changes (e.g., \cite{Kok18}), allowing $\tau_s$ to be directly estimated.  With additional observational constraints on $\rho$, $a$, $Q_g$ and $P$, and with the use of Equation \ref{tau_s}, the  dimensionless moment arm, $k_T$, can be estimated \citep{Jew21b}.  The median value is $k_T$ = 0.007, meaning that only 0.7\% of the outflow momentum is used to torque the nucleus.  Even this tiny fraction is sufficient to quickly modify the spins of small comets.  

\begin{figure}
\epsscale{0.7}
\plotone{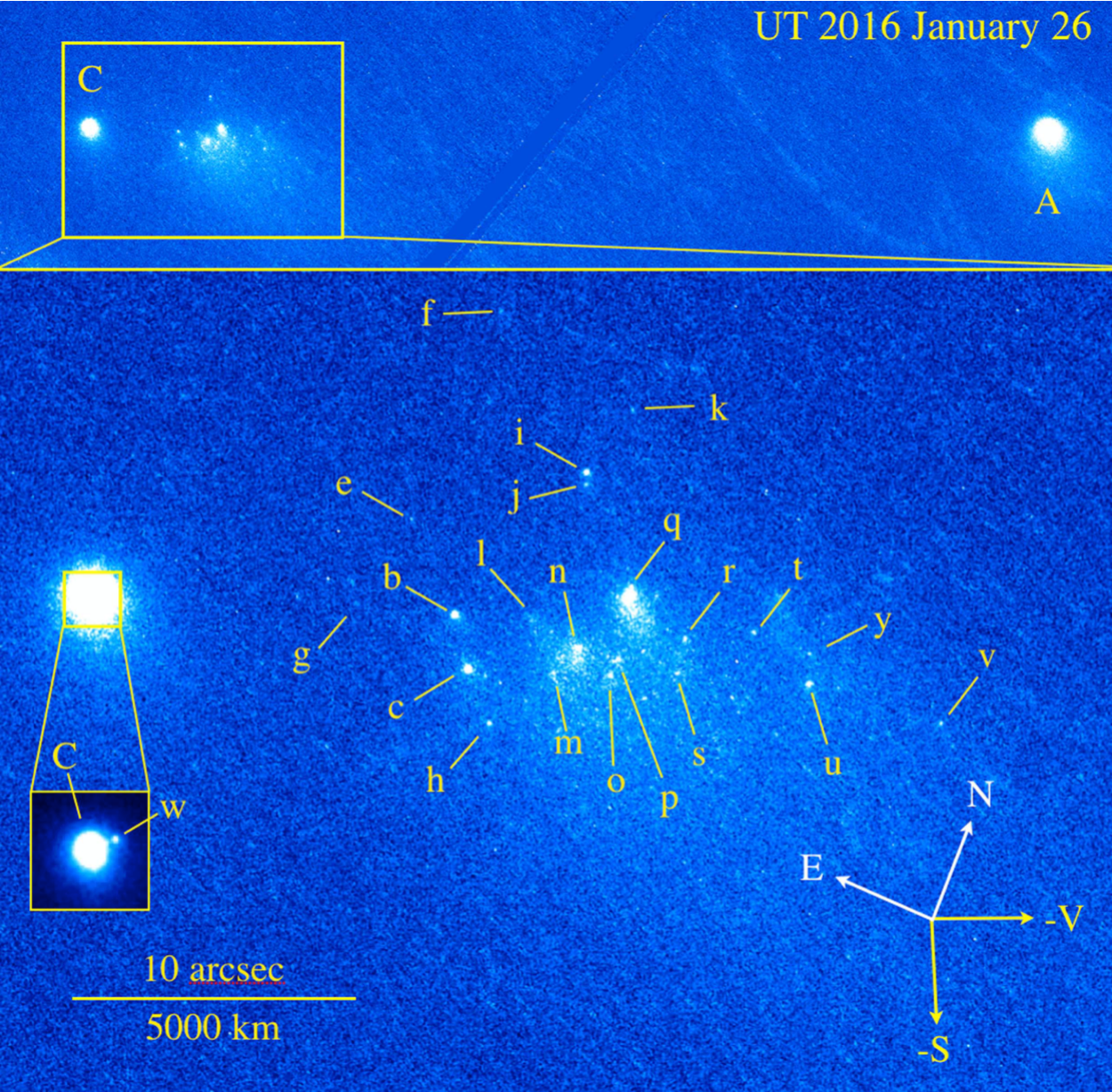}
\caption{Labeled fragments  released from component C of comet 332P/Ikeya-Murakami in 2015 and separating from it at speeds 0.06 to 4 m s$^{-1}$.  C itself rotates with a probable period near 2 hours, suggesting rotational instability as the cause of the release of fragments.  From \cite{Jew16}.
 \label{332P}}
\end{figure}

Measured values  of $\tau_s$ as a function of $a$ are plotted in Figure \ref{comets_spin}.  The relation

\begin{equation}
\tau_s \sim 100  a^2
\label{taufit}
\end{equation}

\noindent with $a$ in km and $\tau_s$ in years, matches the data well \citep{Jew21b}.  (This relation strictly applies to JFCs with perihelia in the 1 to 2 au range). Spin-up by outgassing torques ends with rotational instability and breakup.  The fragmented nucleus of 332P/Ikeya-Murakami at $r_H$ = 1.6 au gives a good example (Figure \ref{332P}).  A cloud of fragments expands from  main nucleus C, whose rotation in about 2 hours suggests rotational instability \citep{Jew16}.

A necessary condition for Equation \ref{tau_s} to remain valid is the persistence of ice at or near the physical surface of the nucleus for timescales $\ge \tau_s$.  Near surface ice persists because the speed with which the ice sublimation surface erodes into the nucleus exceeds the speed with which heat conducts into the interior, causing fresh ice to be continually excavated. The particular problem is that sublimation exposes refractory particles too large to be ejected by gas drag. These should eventually clog the surface, producing a ``rubble mantle'' and inhibiting or shutting down further sublimation.  How this works in detail is not known, even after several years of in-situ investigation of the nucleus of 67P/Churyumov-Gerasimenko by the Rosetta spacecraft \citep{Att23}.  Orbital evolution may play a role, particularly when the perihelion distance migrates to smaller values, leading to higher temperatures and sublimation fluxes.  In the case of the active asteroids, ice is exposed only intermittently, perhaps in response to occasional impacts that clear a surface refractory mantle.  The duty cycle (ratio of the time spent in sublimation to the total elapsed time) is less than $10^{-4}$ or even $10^{-5}$, allowing ice to persist for long times but rendering Equation \ref{taufit} inapplicable to these objects.

The significance of the short timescales indicated by Equation \ref{taufit} is that small nuclei cannot survive long once they reach the vicinity of the Sun.  This may explain the paucity of sub-kilometer comet nuclei relative to power-law extrapolations from larger sizes.  It also complicates any attempt to relate the populations and properties of comets near the Sun with their similarly sized counterparts in the Kuiper belt and Oort cloud source regions.

\subsection{YORP Torque}
\label{YORP2}
Non-spherical bodies warmed by the Sun radiate infrared photons, which carry away momentum and can exert a torque (e.g., \cite{Rub00}).   Here
we offer a simplified treatment that captures the physical essence of the torque, followed by some examples of its application to real bodies. The torque is proportional to the surface area, $a^2$, multiplied by the moment arm, which we write as $k_T^{'} a$, where $0 \le k_T^{'} \le 1$ is another dimensionless constant that must be empirically determined.  Again setting timescale $\tau_{YORP} \sim L/T$ we obtain, 

\begin{equation}
\tau_{YORP} = \frac{16 \pi \rho a^2 c}{15 k_T^{'} P} \left(\frac{r_{au}^2}{S_{\odot}}\right)
\label{tau_YORP}
\end{equation}

\begin{figure}
\epsscale{0.99}
\plotone{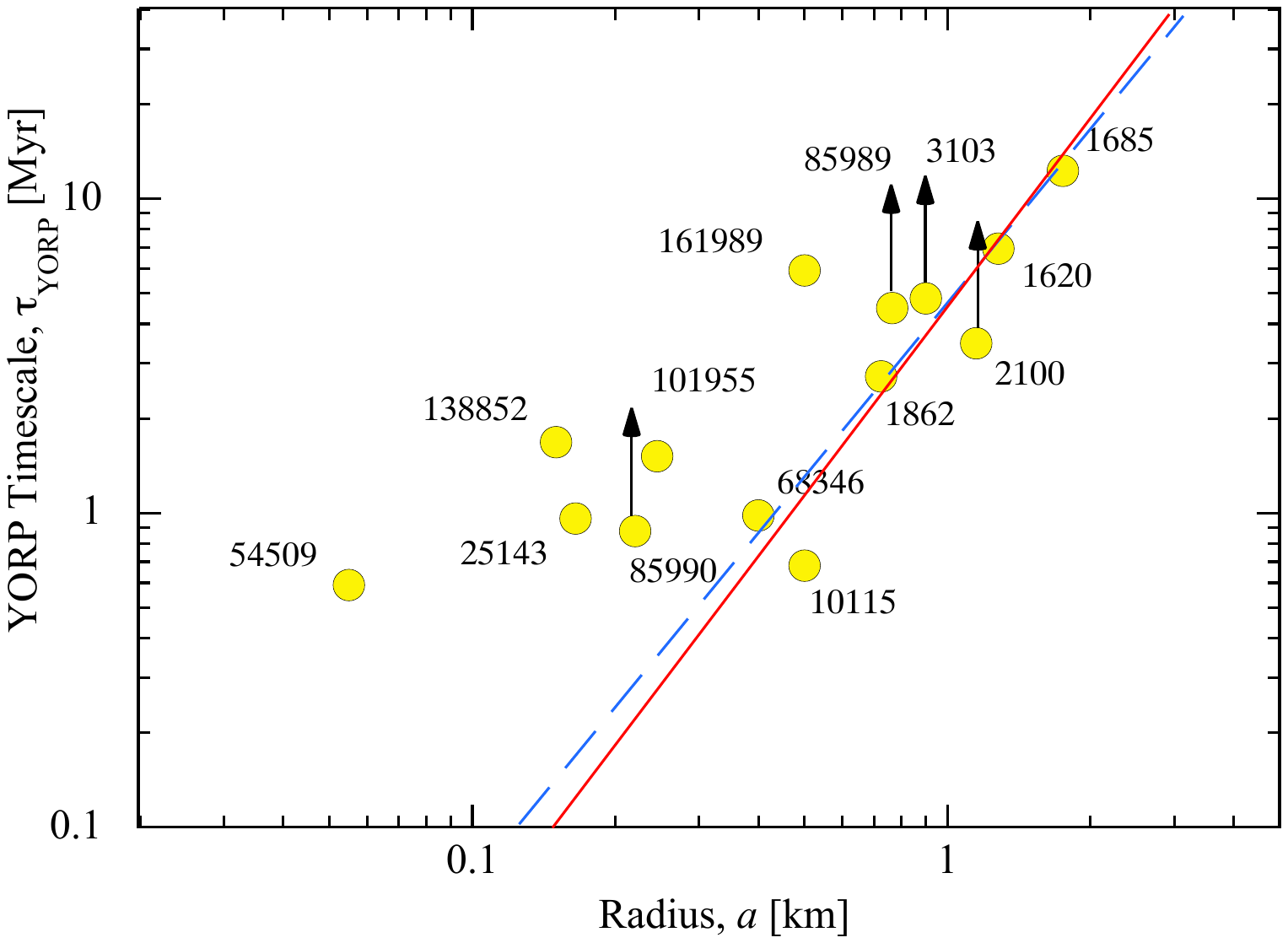}
\caption{Empirical timescale for YORP spin-change for asteroids near 1 au as a function of the asteroid radius [km]. Data with upward pointing arrows represent $3\sigma$ lower limits to $\tau_{YORP}$ based on non-detections of rotational acceleration.  They are not included in the fit.  The dashed blue line shows a weighted least-squares fit  to the detections, $\tau_{YORP} \propto a^{1.87\pm0.04}$.  The solid red line shows Equation \ref{yorpfit} for $r_{au}$ = 1. Data from \cite{Dur24}.
 \label{asteroids_spin}}
\end{figure}

 The  dimensionless moment arm, $k_T^{'}$, is such a sensitive function of the body shape, surface roughness and thermophysical parameters, all of which are unknown for most asteroids, that it cannot in general be calculated.  Instead, we rely on measurements of the few asteroids where changes in the rotation periods can be measured and the other parameters in Equation \ref{tau_YORP} are constrained.   These asteroids are listed in Table \ref{YORP}, where measurements of $a$, $P$ and $dP/dt$ are from the compilation by \cite{Dur24} (see also \cite{Roz13} and references therein). Figure \ref{asteroids_spin} shows the YORP timescale computed from $\tau_{YORP} = P/(dP/dt)$, with $dP/dt =  -(P^2/2\pi) d\omega/dt$ and $d\omega/dt$ from the Table.  $\tau_{YORP}$ is shown as a function of the asteroid radius with each object labeled by its identifying number. The asteroids used in Figure \ref{asteroids_spin} all have semimajor axis $\sim$1 au. 

The Figure clearly shows that $\tau_{YORP}$ and $a$ are correlated, with larger asteroids showing longer YORP timescales.   Adopting the form of Equation \ref{tau_YORP} and the data from Figure \ref{asteroids_spin}, we estimate

\begin{equation}
\tau_{YORP} \sim 4.5~ a^2 r_{au}^2~~\textrm{[Myr]}
\label{yorpfit}
\end{equation}

\noindent where $a$ is the radius in km and $r_{au}$ is the semimajor axis in au.  By this equation, across the asteroid main belt from 2.1 to 3.3 au, a 1 km radius asteroid  would have $\tau_{YORP} \sim$ 20 Myr to 50 Myr.  Setting $\tau_{YORP} = 4,500$ Myr in Equation \ref{yorpfit} and solving for $a$, we find that YORP can influence the spins of main-belt asteroids up to $a \sim$ 10 to 14 km.  This is in broad agreement with the asteroid rotational barrier (Figure \ref{p_vs_r_JPL}) which is prominent for asteroids up to about 10 km in diameter (implying spin-up and break-up for smaller bodies) but less evident beyond about 30 km, where primordial spin is likely preserved (e.g.,  \cite{Pra02}).  It should be noted that the dashed blue line in Figure \ref{asteroids_spin} shows a least-squares fit to the data, weighted by the estimated uncertainties on $\tau_{YORP}$ (but not on $a$).  It gives $\tau_{YORP} \propto a^{1.87\pm0.04}$.  Although the  value of the index is formally (3$\sigma$) smaller than the expected value, $\tau_{YORP} \propto a^2$ from Equation \ref{tau_YORP} (shown in the Figure as a solid red line), the difference is likely not important given the existence of systematic errors in the sample.

A major limitation to the order-of-magnitude estimates of the torque timescales, both for sublimation (Equation \ref{taufit}) and for YORP (Equation \ref{yorpfit}), concerns the stability of the torque.  In comets, the magnitude and direction of the outgassing torque depend on the surface distribution and angular dependence of the outgassing.  As the surface erodes, we expect the spatial and angular distribution of outgassing sources to change, altering the torque.  For most comets, we possess few or no observational constraints on the areal distribution of sources or their evolution.  Indeed, in-situ measurements from 67P/Churyumov-Gerasimenko show the difficulty in modeling this process even given the most detailed data \citep{Att23}.  Likewise, the YORP torque is highly sensitive to the surface shape and texture  and can change in magnitude and even direction in response to minor surface changes \citep{Sta09}. Empirical but indirect evidence for this comes from asteroid (162173) Ryugu ($a \sim$ 0.5 km) which has the characteristic diamond shape indicative of rotational instability but a current rotation period near 7.6 hours.  The rotation of Ryugu may have slowed in response to a changing YORP torque, leaving its equatorial ridge as evidence of its previously rapid spin.   Spin evolution in the presence of a changing torque may be more akin to a random walk process than to the steady change implicit in Equations \ref{taufit} and \ref{yorpfit} (\cite{Bot15}, \cite{Sta15}). The relevant spin-change timescales would then be much longer than estimated here.  Unfortunately, we possess too little information to address this problem with any confidence.

\subsection{Examples:}
\begin{figure}
\epsscale{0.99}
\plotone{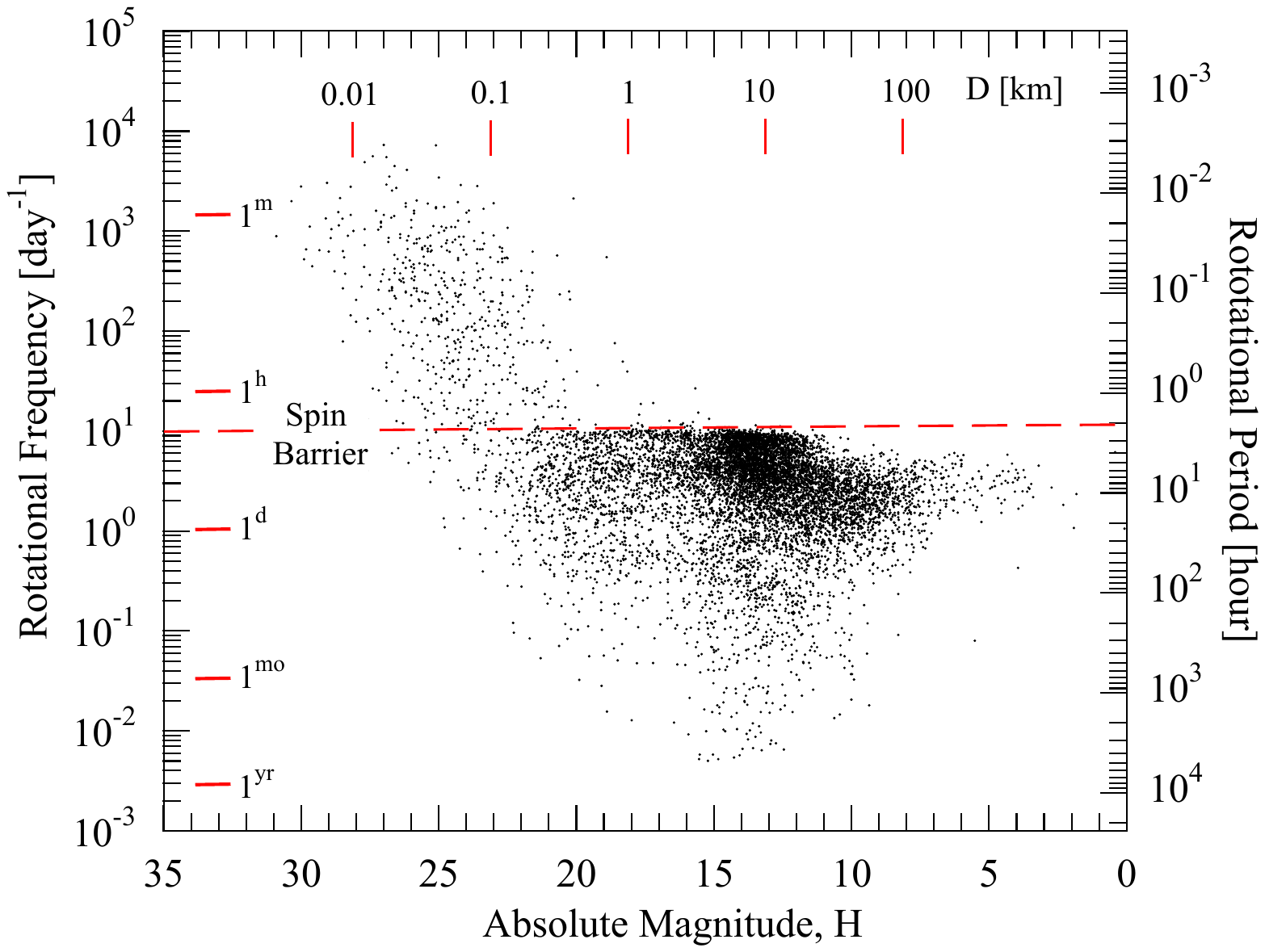}
\caption{Distribution of  asteroid rotational frequencies [rotations day$^{-1}$] as a function of absolute magnitude, $H$. Approximate asteroid diameters  are marked at the top of the figure while their rotation periods are indicated on the right hand axis. The dashed red line marking the ``spin barrier'' at $P^{-1} \sim$ 10 day$^{-1}$ shows that few asteroids with diameter $D \gtrsim$ 0.2 km have periods $<$2.4 hr.    Data from \cite{War09} (updated 2023 October 1 from the F-D Basic file at \url{http://www.MinorPlanet.info/php/lcdb.php}). \label{p_vs_r_JPL}}
\end{figure}

\begin{itemize}

\item \textbf{Asteroid Spin Barrier:} The distribution of  asteroid spin frequencies as a function of absolute magnitude, $H$, is shown in Figure \ref{p_vs_r_JPL}.  $H$ is related to 
the asteroid diameter by $D = 1329\times 10^{-0.2H}/p^{1/2}$, where $p$ is the geometric albedo.  (We assumed $p$ = 0.1 to calculate the diameters shown along the top axis of the figure; in fact, the albedos of a majority of the plotted asteroids have not been measured). Few asteroids larger than $D \sim$  0.2 km  have periods shorter than $\sim$2.4 hour (marked in the figure by a horizontal, dashed red line) whereas, among  smaller asteroids, more rapid rotation is common (\cite{Pra02}, \cite{Ben22}).  This is (admittedly indirect) evidence for spin-up by YORP of a ``bunch of grapes'' internal structure in which  centripetal accelerations in asteroids with periods $\lesssim$2.4 hour result in mass shedding or breakup into smaller, more cohesive component units.  Measured periods of some tiny asteroids are as short as a few seconds \citep{Ben22}.

\item \textbf{Asteroid Reshaping and Pair Formation:} Separate evidence for rotational reshaping of asteroids is provided by the morphologies of some spacecraft-visited asteroids, which show approximate rotational symmetry but with an equatorial skirt consisting of material that has evidently migrated from higher latitudes \citep{Sch15}. New observations of asteroids have also revealed cases of episodic (311P/Panstarrs (2013 P5); \cite{Jew13}) and catastrophic (P/2013 R3 (Catalina-PanSTARRS); \cite{Jew17}) mass loss  that indicate mass shedding and rotational breakup in real-time.  Like 311P, the 2 km radius asteroid (6478) Gault displayed episodic mass loss consistent with mass shedding instability and also has a rotation period (2.55 hour) suggestively close to the rotational barrier \citep{Luu21}.  See Figures \ref{episodic} and \ref{breakups}.  

Extreme end-cases of continued spin-up under the influence of YORP torque include rubble pile disaggregation \citep{Sch18}, which might have been observed in P/2013 R3 (Figure \ref{breakups}), and the formation of asteroid binaries and pairs.  These are independent asteroids with  orbital element similarities that are statistically improbable by chance alone (\cite{Jac11}, \cite{Wal18}).  Asteroid pairs show a systematic relation between the angular frequency of the primary and the secondary/primary mass ratio, such that high mass ratio pairs have distinctly long period primaries \citep{Pra19}.  This is a result of the combined action of primary spin-up by YORP torques and tidal transfer of rotational energy from the primary, needed to expand the orbit of the secondary. Sudden mass transfer events on asteroids should lead to excited rotation; relevant observations are difficult and presently lacking.  

\item \textbf{Spin Alignment:}  It might be expected that collisionally produced asteroid families should have a very broad or even random distribution of spin vectors.  \cite{Sli02} discovered that the spin vectors of the Koronis family asteroids are instead clustered, with obliquities preferentially near 45\degr~and 170\degr.  This clustering was subsequently modeled as a consequence of interaction between gravitational and YORP torques \citep{Vok03}.  Subsequent work showed that the general asteroid obliquity distribution is size dependent \citep{Han11}. Objects of diameter $\lesssim$30 km are more likely to have YORP-aligned obliquities near 0\degr~and 180\degr~(see Figure 7a of \cite{Han11}). The efficacy of alignment  by radiation torques is dependent not just on size but also on many poorly constrained physical and thermophysical parameters  \citep{Gol21}. The timescale for changing the obliquity can be shorter than the timescale for changing the spin \citep{Sta15}.

\begin{figure}
\epsscale{0.7}
\plotone{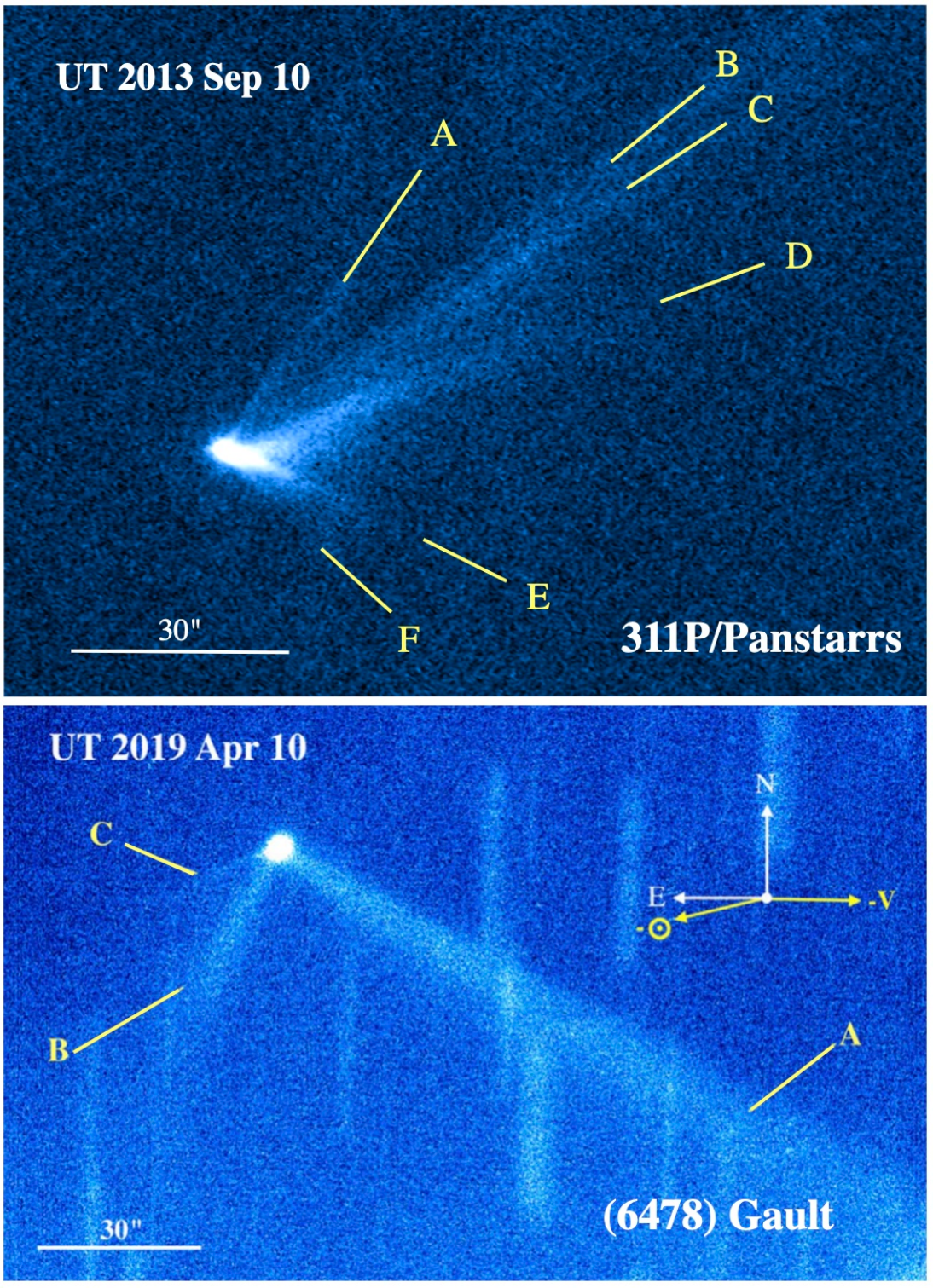}
\caption{Two active asteroids showing episodic ejections likely due to YORP-driven rotational instability.  Each tail corresponds to a particular synchrone; the tail position angle is a measure of the date of ejection.  311P/Panstarrs (top) exhibited nine ejections over about nine months (six visible on 2013 Sep 10 plus three later) \citep{Jew13} while (6478) Gault displayed three tails in 2019 \citep{Luu21}.  
 \label{episodic}}
\end{figure}

\begin{figure}
\epsscale{0.9}
\plotone{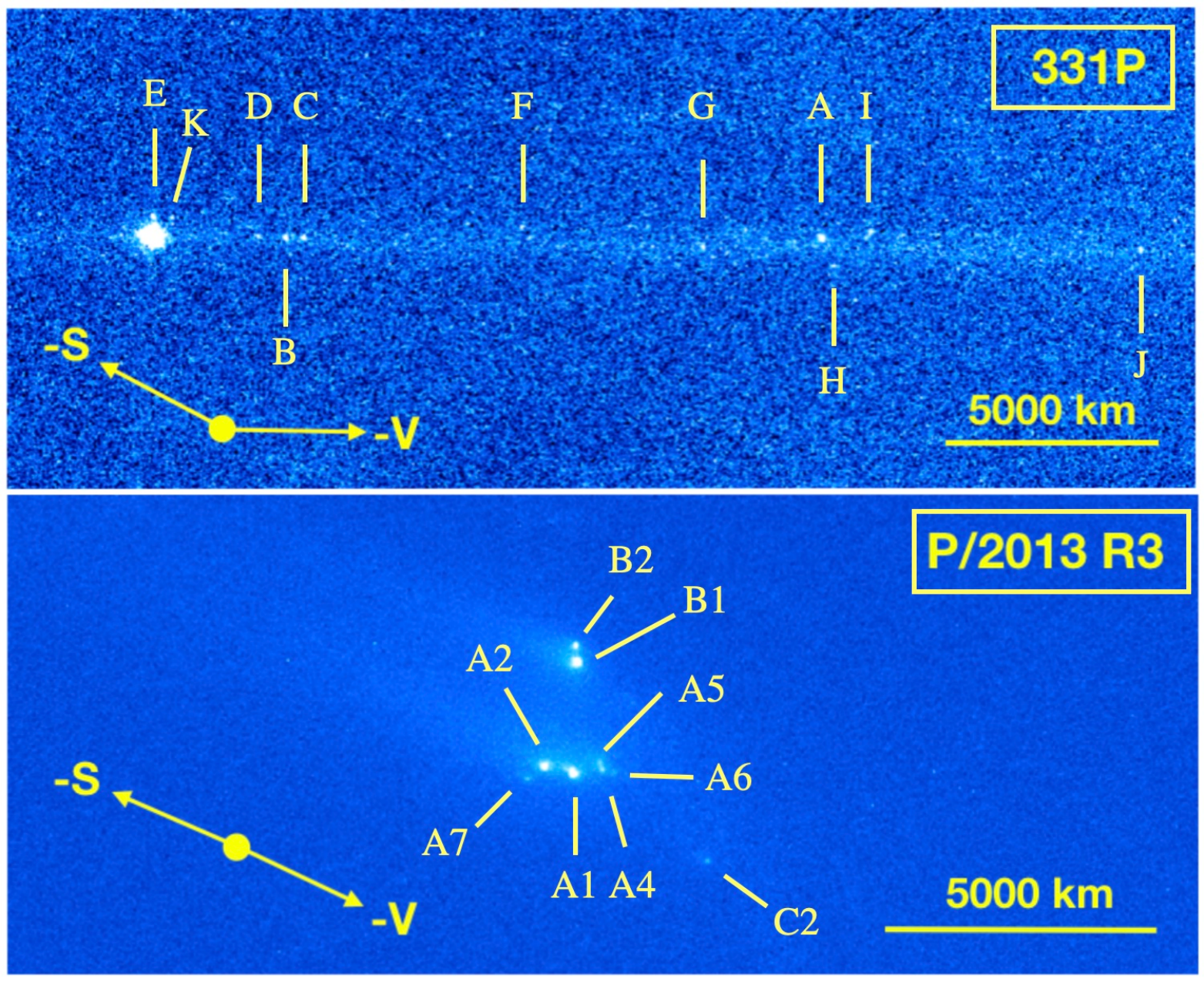}
\caption{Rotationally disrupted asteroids 331P/Gibbs (top, from \cite{Jew21}) and P/2013 R3 (bottom, from \cite{Jew17}).  Major fragments in each object are labelled. The largest of these in 331P is the 100 m scale fragment A, which is also a contact binary.  Sizes of the fragments in P/2013 R3 are less certain because of heavy dust contamination, but estimated to be $\lesssim$100 m.  Both objects are active asteroids, possessing cometary designations but having asteroid orbits \citep{Jew12}. 
 \label{breakups}}
\end{figure}

\item \textbf{Post-Main Sequence Evolution:} As with the Yarkovsky force, the YORP effect will also be magnified as the Sun depletes its core hydrogen and its luminosity surges in the giant phase, by a factor up to $\sim10^{3.5}$ \citep{Vas93}.  Rotational disruption of extrasolar asteroids by YORP spin-up may be an important contributor to debris that  contaminates the photospheres of some white dwarf stars  \citep{Ver14}, where median mass accretion rates are in the range 10$^5$ kg s$^{-1}$ to 10$^6$ kg s$^{-1}$ \citep{Wil24}.

\item \textbf{{Tangential YORP (TYORP):}}
The YORP torque described above is a result of gross deviations from rotational symmetry in the body shapes of asteroids.  This is sometimes referred to as normal YORP or NYORP.  Another radiation-driven torque, known as tangential YORP (TYORP), can be supplied by rocks and other small-scale surface structures \citep{Gol12}.  TYORP is different from NYORP in that it can exist even on a spherical asteroid, provided its surface is littered with rocks of the appropriate scale. Consider a rock of size, $a$, on the surface of an asteroid rotating with period, $P$.  The conduction cooling time of the rock is $\tau_{c} \sim a^2/\kappa$ (Equation \ref{tcool}), where $\kappa$ is the thermal diffusivity.  If  $\tau_c \ll P$,  the rock temperature will quickly equilibrate to the changing insolation through the day.  Its temperatures at sunrise and sunset will be equal.  If $\tau_c \gg P$, the rock temperature will be determined by the average of the day/night insolation, with little temporal variation.  Only when $\tau_c \sim P$ will the rock temperature exhibit substantial diurnal variation, being hotter near local sunset than near local sunrise because of cooling in the night.  This temperature asymmetry can produce a torque with a large moment arm, because radiation and recoil from the warm sun-facing side of a rock at sunset is not balanced by radiation and recoil from an equally warm sun-facing side at sunrise.  The critical rock size is $a_{TY} \sim (P \kappa)^{1/2}$.  For example, with $P$ = 5 hours, $\kappa = 10^{-6}$ m$^2$ s$^{-1}$ (as appropriate for a non-porous, consolidated material), $a_{TY} \sim$ 0.15 m.   Decimeter sized rocks exert particularly strong tangential YORP torques.

The magnitude of TYORP again depends on many unknown properties of the asteroid surface, notably the size and spatial distributions of surface rocks, and the degree to which each rock is separated from its neighbors enough to be thermally independent. But under some circumstances, the magnitude of TYORP may rival that of NYORP \citep{Gol12}.  Moreover, the TYORP and NYORP  torque vectors need not act in the same direction, raising the possibility that one might cancel the other, leading to no net torque on a body that otherwise might be expected to exhibit rotational acceleration.

\item \textbf{{Binary YORP (BYORP):}}
Angular momentum added by photons  can have particularly dramatic dynamical effects on some asteroid binaries \citep{Cuk05}.  By itself, tidal dissipation in close binaries commonly leads to synchronous rotation of the secondary (in which the orbit period and rotation period of the secondary are the same). In the synchronous state,  the secondary can act effectively as an asymmetric extension of the primary, providing a large lever arm for the action of the radiation torque \citep{Cuk05}.  
This torque is known as BYORP (B for binary).  In non-synchronous binaries the effect of BYORP is averaged to zero. The evolution of binary asteroids under the combined action of tidal dissipation and BYORP radiative torque can lead to complicated orbital and spin evolution \citep{Jac11}, with evolutionary timescales as short as $\sim10^4$ years \citep{Ste11}.  If the tidal torque on the secondary is  small compared to the radiative torque  the satellite can escape synchronous rotation and follow its own spin evolution, potentially leading to rotational breakup in-orbit.  The fragments from such a breakup would have short mutual collision times and quickly reaccumulate into a new, rubble pile satellite that repeatedly transforms into a ring before collapsing back into a single body.

\subsection{Summary}
The timescales for the action of water ice sublimation torque ($\tau_S$, Equation \ref{tau_s}) and YORP torque ($\tau_{YORP}$, Equation \ref{tau_YORP}) are compared as functions of heliocentric distance in Figure \ref{yorp_sub}.  In the figure, solid and dashed lines denote assumed object radii of 1 km and 0.1 km, respectively.   The sublimation model was computed assuming $\overline{\cos(\theta)}$ = 1/2 in Equation \ref{sublimation}, which lies in between the high and low temperature limit models in Figure \ref{ices}, and is scaled to timescale $\tau_s$ = 10$^2$ years on a 1 km nucleus at 1 au, in agreement with the data (Figure \ref{comets_spin}).  The figure shows that, for a given object size and distances $\lesssim$5 or 6 au, the timescales for spin change are $\sim$10$^5$ times shorter for sublimation torques than for YORP torques.    Beyond Saturn, the diminishing water ice sublimation rate pushes the sublimation timescale to exceed that from YORP and, in practice, both processes become so slow in the middle and outer solar system as to become largely irrelevant. 
 
\begin{figure}
\epsscale{0.99}
\plotone{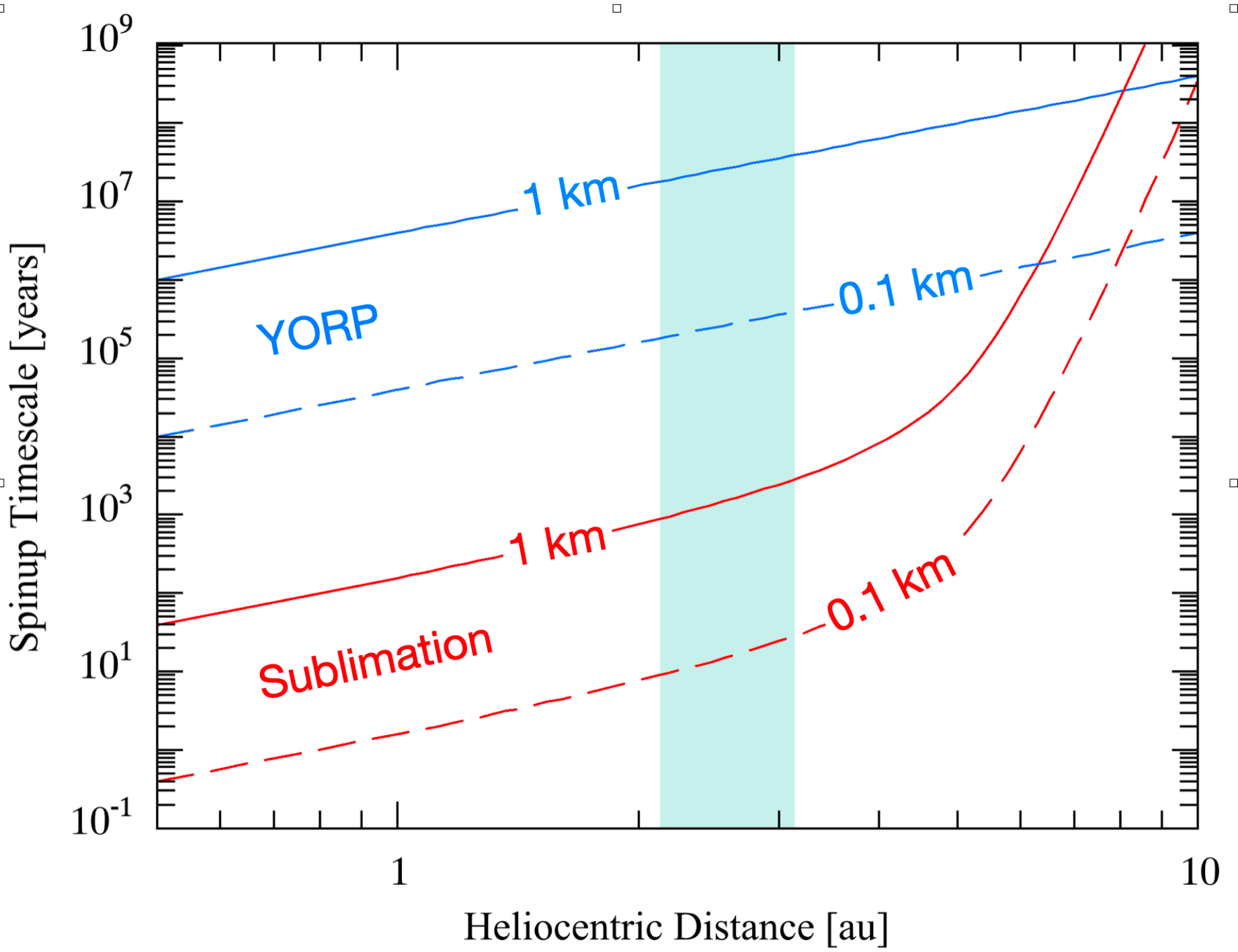}
\caption{The YORP (blue) and water ice sublimation (red)  timescales plotted as a function of $r_H$.  Timescales for objects 1 km radius (solid lines) and 0.1 km in radius (dashed lines) are shown.  The shaded band marks the approximate location of the main asteroid belt, across which sublimation supplies a $\sim10^5$ times stronger torque if near-surface ice is present.
 \label{yorp_sub}}
\end{figure}

\end{itemize}

\clearpage

\renewcommand{\theequation}{A.\arabic{equation}}

\setcounter{equation}{0}

\renewcommand{\theequation}{A.\arabic{equation}}

\setcounter{equation}{0}
\textbf{\appendix{Appendix A: Thermophysics}}
\label{AppendixA}
To gain some physical insight into how the thermophysical parameters of an asteroid might affect Yarkovsky and YORP, we consider the 1-dimensional heat conduction equation

\begin{equation}
\rho c_p\frac{\partial T}{\partial t} = k \frac{\partial^2T}{\partial z^2},
\label{diffusion}
\end{equation}

\noindent where $T$ is the temperature, $k$ is the thermal conductivity, $\rho$ is  the bulk density and $c_p$ is the specific heat capacity of the asteroid surface materials.   Dimensionalizing this equation gives the timescale for heat to conduct over a distance $\ell$ as

\begin{equation}
\tau_C \sim \frac{\ell^2}{\kappa}~~~~~~\textrm{where}~~~~~~
\kappa = \frac{k}{\rho c_p}
\label{tcool}
\end{equation}

\noindent and $\kappa$ [m$^2$ s$^{-1}$] is the thermal diffusivity.\footnote{Diffusivity appears directly in Equation \ref{diffusion} and is the natural measure of thermal response of a material through Equation \ref{tcool}. In planetary science, the use of thermal inertia is instead widely preferred.  Thermal inertia is defined by $I = (k \rho c_p)^{1/2}$, which has the somewhat uncomfortable units [J m$^{-2}$ s$^{-1/2}$ K$^{-1}]$.  Solid dielectrics (e.g., non-porous rocks) have $I \sim 10^3$ J m$^{-2}$ s$^{-1/2}$ K$^{-1}$ while the finest dust, as found in the regoliths of small outer solar system bodies, has $I \sim$ 1 to 10 J m$^{-2}$ s$^{-1/2}$ K$^{-1}$ \citep{Fer18}. Asteroid inertias in the range 10 $\le I \le$ 100 J m$^{-2}$ s$^{-1/2}$ K$^{-1}$ are common \citep{Mac21}.} For a body with rotation period $P$, we set $\tau_C = P$ and use Equation \ref{tcool} to find the distance over which heat can conduct in one rotation period as $\ell = (\kappa P)^{1/2}$, which defines the diurnal thermal skin depth on the asteroid.  An exact solution of Equation \ref{diffusion} would show that $T$ varies with depth as a damped sinusoid, with $(\kappa P)^{1/2}$ being the e-folding length scale of the damping, but our order of magnitude approximation is sufficient here.  On a spherical body with radius $a$, the heat contained within the skin depth, $\ell$, is

\begin{equation}
H = 4 \pi a^2 \rho (\kappa P)^{1/2} c_p T
\label{H}
\end{equation}

\noindent where $c_p$ [J K$^{-1}$ kg$^{-1}$] is the specific heat capacity of the shell.   Loss of heat by radiation into space occurs at the rate

\begin{equation}
\frac{dH}{dt} = 4 \pi a^2 \sigma T^4
\label{dHbdt}
\end{equation}

\noindent where $\sigma = 5.67\times10^{-8}$ W m$^{-2}$ K$^{-4}$ is the Stefan-Boltzmann radiation constant and we assume emissivity $\varepsilon$ = 1.  The order-of-magnitude cooling timescale is $\tau_c \sim H/(dH/dt)$, so that we can use Equations \ref{H} and \ref{dHbdt} to write the thermal parameter, $\Theta = \tau_c/P$  (c.f., \cite{Spe89})

\begin{equation}
\Theta = \frac{\rho c_p}{\sigma T^3}\left(\frac{\kappa}{P}\right)^{1/2}.
\label{tau_over_P}
\end{equation}

\noindent Asteroids with very rapid rotation ($\Theta \gg$ 1)  retain their heat through the night and should remain nearly longitudinally isothermal, producing little net diurnal Yarkovsky force.  Similarly, asteroids with $\Theta \ll$ 1 will experience little diurnal Yarkovsky drift because rotation is too slow to carry the diurnal temperature maximum away from the midday meridian. Intuitively, on the other hand, asteroids with $\Theta \lesssim$ 1 should retain heat into the afternoon but lose it by the morning and so will experience a net force from asymmetric radiation\footnote{With more effort than is warranted here, it can be shown that the effect is proportional to $\Theta/(1 + 2\Theta + 2\Theta^2)$, a function that is maximized for $\Theta \sim$ 0.7 \citep{Rub95}.}.     

As a rough example, consider an asteroid with a very porous surface regolith having $\kappa$ = 10$^{-9}$ m$^2$ s$^{-1}$ and with nominal $\rho = 10^3$ kg m$^{-3}$, $c_p = 10^3$ J K$^{-1}$ kg$^{-1}$, $P = 2\times10^4$ s (i.e., about 6 hours) and orbiting near 1 au, where $T \sim$ 300 K.  Substitution  gives skin depth $\ell \sim$ 5 mm, $\Theta \sim$ 0.15, and the peak heat of noon will be retained for about 1/6th of a rotation, corresponding to a lag angle $\theta \sim$60\degr. The Yarkovsky force on a small asteroid with these parameters is likely to be significant.   By comparison, an asteroid having the same properties but consisting of solid rock, for which $\kappa = 10^{-6}$ m$^2$ s$^{-1}$ would have  $\ell \sim$ 15 cm, $\Theta \sim$ 4.5 and, by virtue of taking several rotations in order to cool, would  experience minimal diurnal temperature variation and reduced Yarkovsky acceleration.  

These simple considerations are only illustrative. Real asteroids are complicated and usually poorly characterized, with irregular shapes and boulder-strewn, variegated surfaces. On such bodies the distribution of surface temperature and the magnitudes of the Yarkovsky and YORP effects cannot be accurately calculated, only worked-out after the fact by measuring radial drift and changes in the rotation period.  Still, the description given above provides a useful framework for understanding the action of radiative non-gravitational effects.


\acknowledgments
I thank Marco Fenucci for providing a digital table of his Yarkovsky drift measurements, Pedro Lacerda, Jane Luu, Joe Masiero, Darryl Seligman, David Vokrouhlicky, Emerson Whittaker and an anonymous referee for comments.

\clearpage

\begin{deluxetable}{lcrrrr}
\tabletypesize{\scriptsize}
\tablecaption{YORP Timescales\tablenotemark{a}
\label{YORP}}
\tablewidth{0pt}
\tablehead{\colhead{Object}   &  \colhead{$D$\tablenotemark{b}} &  \colhead{$P$\tablenotemark{c}} & \colhead{$d\omega/dt$\tablenotemark{d}}  & \colhead{$\tau$\tablenotemark{e}}   }

\startdata
(1620) Geographos &  2.56 & 5.22 & 1.14$\pm$0.03 & 6.9$\pm$0.2  \\

(1685) Toro & 3.5 & 10.20 & 0.33$\pm$0.03  & 12.3$\pm$1.0  \\

(1862) Apollo &  1.55 & 3.07  & 4.94$\pm$0.09 & 2.7$\pm$0.05 \\

(2100) Ra-Shalom   &  2.30 & 19.82  & $<$0.6 & $>$3.5 \\

(3103) Eger &  1.78 & 5.71  & $<$1.5 & $>$4.8 \\  

(10115) 1992 SK & 1.0 & 7.32 & 8.3$\pm$0.6 & 0.68$\pm$0.05  \\

(25143) Itokawa &  0.32 & 12.13  & 3.54$\pm$0.38 & 1.0$\pm$0.1 \\

(54509) YORP &  0.11 & 0.20  & 350$\pm$35 & 0.59$\pm$0.05 & \\

(85989) 1999 JD6 &  1.53 & 7.66 & $<$1.2  & $>$4.5  \\ 

(85990) 1999 JV6 & 0.44 & 6.54 & $<$7.2 & $>$0.9  \\

(68346) 2001 KZ66 & 0.80 & 4.99 & 8.43$\pm$0.69 & 1.0$\pm$0.1&  \\

(101955) Bennu & 0.49 & 4.30 & 6.34$\pm$0.91 & 1.5$\pm$0.2  \\

(138852) 2000 WN10 &  0.3 & 4.46  & 5.5$\pm$0.7 & 1.7$\pm$0.2 \\

(161989) Cacus &  1.0 & 3.76 & 1.86$\pm$0.09 & 5.9$\pm$0.3 \\
\enddata

\tablenotetext{a}{Asteroid data from \cite{Dur24}}
\tablenotetext{b}{Diameter in km} 
\tablenotetext{c}{Rotation period in hours}
\tablenotetext{d}{Rate of change of the angular frequency $\times10^{-8}$ [radian day$^{-1}$]. Values from \cite{Dur24} which are statistically insignificant are listed as 3$\sigma$ upper limits. }
\tablenotetext{e}{YORP time, Myr}
\end{deluxetable}

\clearpage

\end{document}